\documentclass[a4paper,fleqn,usenatbib]{mnras}
\usepackage[T1]{fontenc}
\usepackage{ae,aecompl}

\usepackage{bm}

\usepackage{graphicx}	
\usepackage{amsmath}	
\usepackage{amssymb}	

\usepackage{color}   

\usepackage{amsfonts}
\usepackage{bbm}
\usepackage{times}
\usepackage{rotating}
\usepackage{array}
\usepackage{epsfig}
\usepackage{textcomp}

\newcommand{\HI}{$\mathrm{HI}$\,} 
\newcommand{\Tsys}{T_{\rm sys}}      
\def\uv{\mathrm{(u,v)}}

\newcommand{\needcitation}{  }
\definecolor{orange}{rgb}{0.8,0.4,0}
\newcommand{\rzcheckdone}[1]{ }
\newcommand{\rzcheck}[1]{  }
\newcommand{\rzrefreq}{}
\newcommand{\rzrefreqa}[1]{#1}
\newcommand{\addarxiv}[1]{#1}

\newcommand{\RefEq}[1]{Eq.~\ref{#1}}
\renewcommand{\vec}[1]{\bm{#1}}

\newcommand{\Mpc}{\ensuremath{\,{\rm Mpc}}}

\title{Sky reconstruction from transit visibilities: \\
PAON-4 and Tianlai Dish Array}   
 
\author[J. Zhang et al.]
{Jiao Zhang$^{1,2,3}$, 
Reza Ansari$^{2}$ \thanks{E-mail:ansari@lal.in2p3.fr},
Xuelei Chen$^{1,3,4}$,
Jean-Eric Campagne$^{2}$, 
\newauthor{
Christophe Magneville$^{5}$, 
and Fengquan Wu$^{1}$} \\
$^1$Key Laboratory of Computational Astrophysics, National Astronomical
Observatories, Chinese Academy of Sciences, Beijing 100012, China\\
$^2$Universit\'e Paris-Sud, LAL, UMR 8607, F-91898 Orsay Cedex, France $\&$ CNRS/IN2P3, F-91405 Orsay, France\\
$^3$University of Chinese Academy of Sciences, Beijing 100049, China\\
$^4$Centre for High Energy Physics, Peking University, Beijing 100871, China\\
$^5$CEA, DSM/IRFU, Centre d'Etudes de Saclay, F-91191 Gif-sur-Yvette, France\\
}

\date{Accepted XXX. Received YYY; in original form ZZZ}
\pubyear{2016}

\begin{document}
\label{firstpage}
\pagerange{\pageref{firstpage}--\pageref{lastpage}}
\maketitle

\begin{abstract}
{\rzrefreq The spherical harmonics $m$-mode decomposition is a powerful sky map reconstruction 
method suitable for radio interferometers operating in transit mode. It can be 
applied to various configurations, including dish arrays and cylinders. } 
We describe the computation of the instrument response function,
the \rzrefreqa{point spread function (PSF),} transfer function, the noise covariance matrix and noise power spectrum.
\rzrefreqa{The analysis in this paper is focused on dish arrays operating in transit mode. }
We show that arrays with regular spacing have more pronounced side lobes as well as structures 
in their noise power spectrum, compared to arrays with irregular spacing, specially
in the north-south direction. A good knowledge of the noise power spectrum $C^{\mathrm{noise}}(\ell)$ is essential 
for intensity mapping experiments as non uniform $C^{\mathrm{noise}}(\ell)$ is a potential 
problem for the measurement of the \HI  power spectrum. 
Different configurations have been studied to optimise the PAON-4 and Tianlai dish array layouts. 
We present their expected performance  \rzrefreqa{ and their sensitivities to the 21--cm emission of the Milky Way 
and local extragalactic \HI clumps.}
\end{abstract}



\section{Introduction}

Measurement of the neutral hydrogen (\HI) distribution through its 21--cm line radiation is a powerful  
method for  studying the statistical properties of Large Scale Structure (LSS) in the Universe, 
complementary to optical surveys.
However, given the very faint radio brightness of typical \HI clumps, detection of individual galaxies in 21--cm 
at cosmological distances ($z \gtrsim 1$) requires very large collecting areas, around $\sim \mathrm{km^2}$. Moreover, extracting 
cosmological information from LSS requires the observation of large volumes of universe 
{\rzrefreq to probe long wavelength modes with sufficient precision in order to be competitive with the optical galaxy surveys}. 
In recent years, the intensity mapping technique has been suggested as an 
efficient and economical way to map large volumes of the universe using the \HI ~21--cm emission. Such 
cosmological surveys would be especially suitable for late time cosmological studies ($z \lesssim 3$), in particular 
to constrain dark energy through {\rzrefreq the Baryon Acoustic Oscillations (BAO) and 
Redshift Space Distortions (RSD) measurements} \citep{2006astro.ph..6104P,2008PhRvL.100i1303C,ansari.IM.2008,2012A&A...540A.129A,2010ApJ...721..164S}. 
In this scheme, the integrated radio emission of many \HI clumps in cells of $\sim 10^3 \, \mathrm{Mpc^3}$ is measured
without detection of individual galaxies. Large wide-field radio telescopes, with an angular resolution
of a fraction of a degree and a frequency resolution of $\lesssim  1 \mathrm{MHz}$ and sensitivities of $\lesssim 1 \, \mathrm{mK}$
per resolution element would be needed to observe the LSS, especially the BAO features.

{\rzrefreq  Several groups throughout the world  are aiming to carrying such surveys. A number of projects with 
single dishes (possibly equipped with multi-beam receivers) and interferometer arrays have been proposed. 
Single dish intensity mapping surveys have been carried out on existing telescopes such as 
the Green Bank Telescope (GBT) \citep{2010Natur.466..463C,2013MNRAS.434L..46S,2013ApJ...763L..20M} 
and construction of dedicated instruments are being planned such as the
BINGO (BAO from Integrated Neutral Gas Observations) project which is 
a single dish radio telescope equipped with an array of feeds in the focal plane \citep{2013MNRAS.434.1239B,dickison.bingo.2014}. 
The interferometer arrays include  CHIME (Canadian Hydrogen Mapping Experiment) in Canada which is currently 
in the final stages of the construction of  5 large cylindrical reflectors \citep{2014SPIE.9145E..22B} and the Tianlai (Chinese for "heavenly sound") 
project in China which has just completed the construction of both a cylinder array and a dish array pathfinders 
in 2015 \citep{2012IJMPS..12..256C} and  the HIRAX (Hydrogen Intensity and Real-time analysis eXperiment)
\footnote{\tt http ://www.acru.ukzn.ac.za/ cosmosafari/wp-content/uploads/2014/08/Sievers.pdf} 
project in South Africa which plans to build a large array of relatively small dishes ($D \sim 6$~m).  Intensity mapping survey 
is also being considered for the upcoming Square Kilometre Array (SKA) mid-frequency dish array, both as an interferometer array
and as a collection of single dishes \citep{2015MNRAS.450.2251Y}. 
The expected results in cosmology from the on-going or projected intensity mapping experiments
are reviewed in \citet{2015ApJ...803...21B}.      }

{\rzrefreq The challenges for the instrument design and data analysis of these experiments are similar to those 
encountered in 21--cm experiments designed to observe the epoch of reionization (EOR) such as 21CMA (21Centimetre Array,
\citealt{2004astro.ph..4083P,2016arXiv160206624Z}), LOFAR (Low Frequency Array, \citealt{2013A&A...556A.2V}), 
MWA (Murchinson Wide-field Array, \citealt{2009IEEEP..97.1497L}, 
and HERA (Hydrogen Epoch of Reionization  Array, \citealt{2015PhRvD..91b3002D}). 
Information on the field of 21--cm cosmology as well as these experiments can be found in a number of excellent reviews such as 
\citet{2006PhR...433..181F,2010ARA&A..48..127M,2012RPPh...75h6901P,2013ASSL..396...45Z}. }


For the intensity mapping surveys, the interferometer arrays operating in transit mode seems to be a natural choice. The 
antennas of such an array are fixed on the ground during observation with the antennae axis in the meridian plane.
For dish arrays, the instantaneous field of view is a small circular patch ($\sim~10~\mathrm{deg^2}$) on the sky
while it is a narrow ($1-2~\mathrm{deg}$) strip along the meridian for cylinders. 
As the Earth rotates, different areas of the sky pass through the field of view. As the telescope does not
need to track the celestial target, the mechanical structure of the telescope is very simple:
it is either fixed on the ground or requires only occasional adjustments in declination for dish arrays. 

{\rzrefreq 
The ultimate goal of a 21--cm intensity mapping experiment is to make precision measurement of the 
cosmological 21--cm 
power spectrum. This is however a very challenging task because of the presence of strong foregrounds astrophysical 
emissions as well as instrument and environmental noise which are both a few orders of magnitude higher than the 
21--cm signal. 
Complex data processing procedures are required and can generally be decomposed into several steps: calibration, map making 
and foreground subtraction. The map making procedure discussed in this paper is a major building block of the processing 
pipeline and determines the instrument and the survey response to both cosmological signal and the foregrounds. 
}

{\rzrefreq  
The observational data from either single dish or interferometer array observations 
are to a good approximation  linearly related to the sky temperature distribution. The map making problem can then be regarded as 
the inverse problem. A number of methods were developed to solve similar problems in cosmic microwave background (CMB) anisotropy 
experiments, read for instance \citep{1997PhRvD..56.4514T} for a review. Broadly speaking, theses methods are applicable 
to 21--cm experiments  with a number of specificities for interferometric arrays. For instance, the map making from visibilities 
for MWA is discussed in  \citep{2012ApJ...759...17S} and the method envisaged for Hydrogen 
Epoch of Reionization Array (HERA) project is presented in \citep{2015PhRvD..91b3002D}. 
An alternative method which do not use visibilities but apply spatial Fourier transform to individual feed signals in order to form beams 
and handles the map making for a regular or semi regular array with very large number of small antennae  was discussed in \citet{2009PhRvD..79h3530T,2010PhRvD..82j3501T}.
}

In this paper we present the method for making maps of the sky from the transit observations 
made with such interferometer arrays and discuss the impact of array configurations on the properties 
of the reconstructed maps.  {\rzrefreq We shall limit ourself to the case of dish arrays though the method is also applicable to 
cylinder arrays which is going to be discussed in a subsequent paper. } 
In Sec.~\ref{sec-arrayintro},  the PAON-4 \rzrefreqa{(PAraboles \`a l'Observatoires de Nan\c{c}ay)}  
and Tianlai Dish Array are briefly introduced.
Section \ref{sec-transitmapmaking} presents an overview of the map making algorithm from 
full east-west transit (24 hours) visibilities or interferometric observations.  
Section \ref{sec-paon4case} discusses the application of the method to the PAON-4 telescope which is a 4 antennae test interferometer and the optimisation of the antennae configuration.  We present the expected beam shapes and the noise power spectrum for PAON-4 compared with a regular $2\times2$ array and a single dish telescope. Section \ref{sec-tianlaicase} present the comparison of beam and noise power spectrum for several array configurations that we have considered for the Tianlai 16-dish pathfinder array
\rzrefreqa{as well as a short discussion of array sensitivities to extragalactic and cosmological 21--cm signals.} 
The conclusions and future work plans are presented in section \ref{sec-conclusions}. 
\addarxiv{The map making code itself and associated tools are presented in appendix \ref{sec-mapmakingcode}. 
The extension of the method to the polarisation is presented in appendix \ref{sec-polarisedmapmaking}  }

\section{The PAON-4 and Tianlai Dish Arrays}
\label{sec-arrayintro}

\subsection{The PAON-4 array}
{\rzrefreq
The PAON-4 array is a small wide band test interferometer (L-band, 1250-1500 MHz) featuring four 5-metre diameter antenna, 
installed at the Nançay radio observatory in France ($47^\circ 22' 55.1"$N, $2^\circ 11' 58.7"$E). 
PAON-4 has been designed and built within the BAORadio project in France \citep{2012CRPhy..13...46A}. 
It has a total collection area of $\sim 75\, \mathrm{m^2}$ and 4 dual polarisation receivers. The dish pointing can be changed in 
declination through computer controlled electric jacks. }
%
The 36 visibilities (8 auto-correlations and 28 cross-correlations) are computed by the BAORadio electronic-acquisition system and 
written to disk with $\sim~1~\mathrm{second}$ time resolution. Tests observations with PAON-4 started in spring 2015 
with the aim of evaluating the use of small dish arrays for intensity mapping and  developing the calibration and map-making 
procedures for such instruments.

The physical {\rzrefreq diameter} of the PAON-4 dishes are $D=5$~m. We model the primary beam of the dish+feed 
as \citep{1999prop.book.....B}, 
\begin{equation}
D(\gamma) \propto \frac{2 \, J_1[\pi (D_{\rm eff} / \lambda) \, \sin \gamma ]}{\pi (D_{\rm eff} / \lambda) \, \sin \gamma} , 
\end{equation}
where $\gamma$ is the angle with respect to the reflector symmetry axis, 
$J_1(x)$ is the first order Bessel function, $\lambda$ the wavelength and  $D_{\rm eff} $ is the effective dish 
diameter illuminating the feed. Based on test observations, 
we have used an efficiency factor $\eta=0.9$ for PAON-4, yielding an effective dish diameter 
$D_{\rm eff} = \eta \, D = 4.5$ m. The single dish first null beam width (FNBM) is 
around $1.22\lambda/D_{\rm eff} \sim 3.25^\circ$ at $1420 \mathrm{MHz}$. 


We studied a number of antenna arrangements before finally chose the PAON-4  configurations. 
In this paper, we shall compare the results for the following configurations. 

\begin{figure*}
\centering
\includegraphics [width=0.8\textwidth] {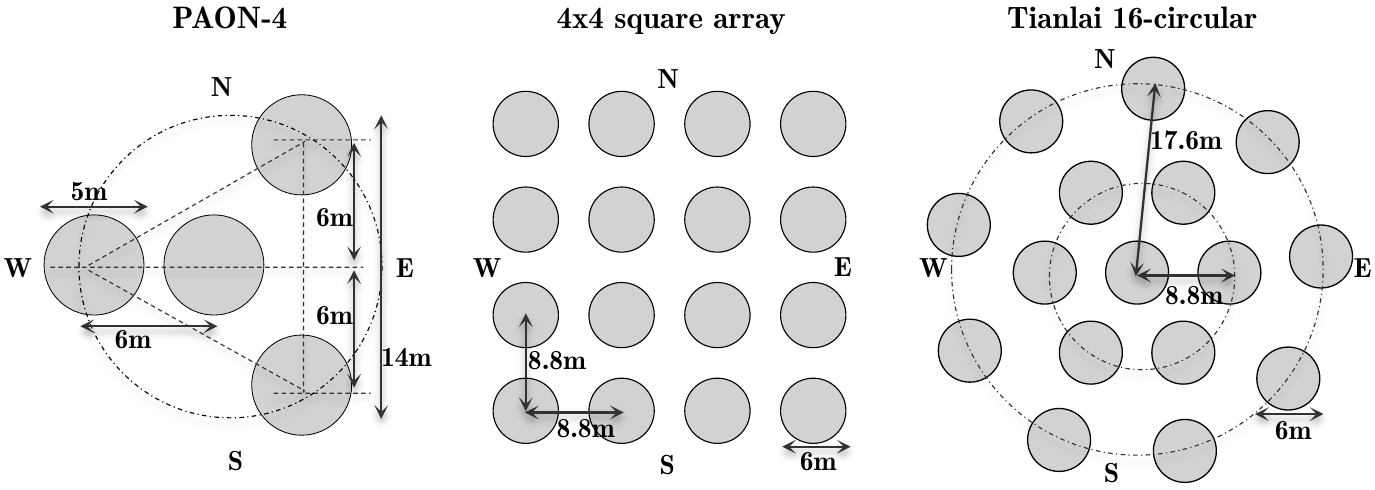}
\caption{The PAON-4 interferometer configuration (left), 
the  regular $4 \times 4$ array (centre) and the 
circular Tianlai array configuration with 16 dishes (right).}
\label{fig-paon4-tianlai-config} 
\end{figure*}

(a) The PAON-4 configuration (shown in the left panel of Fig.~\ref{fig-paon4-tianlai-config}).
Three dishes are arranged at the vertices 
of an equilateral triangle with 12 m sides, one of its side is along the exact North-South line.  
{\rzrefreq The fourth dish is inside the triangle, 6 metres away from the west vertex along the EW baseline.
In addition to the auto-correlation signals, the PAON-4 configuration has 6 different baselines without any redundancy. }

(b) The $2\times 2$ array. The four dishes are arranged into a $2 \times 2$ regular array. 
The sides of the square are aligned with the north-south (NS) and east-west (EW) directions, 
with antennae centres separated by $d_{\rm sep}$. 
{\rzrefreq This configuration has 4 different baselines, in addition to the autocorrelation beam. 
Actually we shall consider two such arrays, a non-compact one with $d_{\rm sep}=14$m (b1) 
and a compact one with $d_{\rm sep}=7m$ (b2). The (b1) configuration 
is discussed first to help understand  the mathematical tools introduced in Sec.\ref{sec-transitmapmaking}, and 
to illustrate the stronger mode mixing (frequency dependent beam) introduced by sparse arrays, due to 
incomplete angular frequency $\uv$ or $(\ell, m)$ planes. The configuration (b2) will be compared to the PAON-4 case. 
}




For PAON-4 we consider a drift scan survey of a full east-west strip of sky centred 
at the PAON-4 latitude ($\delta \simeq 47^\circ$) over a 6 month period. 
The survey will be composed of 25 constant declination scans from $+35^\circ 23'$ to $+59^\circ 23'$, each shifted by 
1 degree in declination. At the end of the survey PAON-4 should be able to provide  sky maps over $\sim 5000~\mathrm{deg^2}$ 
of sky and $\sim 200 \, \mathrm{MHz}$ bandwidth.


\subsection{The Tianlai Dish Array}
{\rzrefreq The Tianlai project is a 21--cm intensity mapping experiment 
aimed at surveying the large scale structure and use its BAO features 
to constrain dark energy models\citep{2012IJMPS..12..256C}. 
The current experiment is a pathfinder for testing the basic principles and key technologies, 
located at a radio quiet site ($44^\circ 10' 47''$ N, $91^\circ 43' 36''$ E) in Hongliuxia, Balikun County, Xinjiang, 
China \citep{2015IAUGA..2252187C}.  The pathfinder includes a cylinder array as well as a dish array.
The discussion of the cylinder array and forecast of its capability can be found in \citet{2015ApJ...798...40X}.
In addition to the cylinder array, the Tianlai pathfinder also includes a dish array with 16 dishes of 6 metres
diameter. These dishes are equipped with electronically controlled motor drives in the altitude-azimuth mount,
which allows the dishes to be pointed to almost any desirable directions above the horizon. 
However, the regular observation mode we envisage for the dish array is to point the 
dishes along the meridian with a common elevation (declination), and make drift scan observations. 
We shall assume that similar to the PAON-4 case,  $\eta=0.9$,  so $D_{\rm eff}=5.4$m, 
and the primary beam (FNBW) for each dish is $2.73^\circ$ at 1420 MHz. Each dish has a dual polarisation 
receiver, which is tunable within the range of 500-1500 MHz with a replaceable bandpass filter 
of 100 MHz bandwidth. The visibilities (32 autocorrelations and 448 cross-correlation) for dish arrays 
are computed by the data-acquisition (DAQ) system and saved in hard drives. 
 The construction of both the cylinder and the dish pathfinder arrays have been completed at the end of 2015, 
the two arrays are now undergoing the commissioning process. }
 
Several antennae layouts for the Tianlai 16-dish array have been investigated, 
but we will focus here on the following two configurations 
(see  the centre and right panels of Fig.\ref{fig-paon4-tianlai-config}): 

(a) \textcolor{blue}{Square Array}(Centre panel of Fig.\ref{fig-paon4-tianlai-config}). 
The antennas are positioned on the grid points of a $4 \times 4 $ square, with the sides 
of the square aligned with the EW and NS directions, and separation for the nearest neighbour grid points 
to be $d_{\rm sep}=8.8$m (in sec.\ref{sec-tianlaicase}  we shall  discuss the choice of $d_{\rm sep}$ in more details). 
The Square Array have large number of redundant  baselines, the total number of independent baselines is only 24 
for each declination pointing. 

(b) Circular Array(Right panel of Fig.\ref{fig-paon4-tianlai-config}). 
This is the configuration we have actually adopted for the current Tianlai dish pathfinder array. 
One antenna is positioned at the centre, the remaining 15 antennas are arranged in two concentric circles
around it. It is well known that the baselines of the circular array configurations are quite independent 
and have an overall wider coverage on the angular frequency $\uv$ plane \citep{2001isra.book.....T}.
We have studied a number of circular arrangements, with or without centre antenna, with one or two 
concentric rings, with different alignment between the inner and outer rings. 
From the investigations, we found that there is no significant difference in terms of map reconstruction performance
between the different circular configurations that were studied. 
The final configuration we chose has 6 and 9 dishes in the inner and outer rings respectively, with radius given 
by the minimal separation of $d_{\rm sep}=8.8$m and $2 d_{\rm sep}$ respectively. The inner ring is 
symmetric with respect to the NS direction, 
\rzrefreqa{with antennae placed every $60^\circ$, starting at $0^\circ$ azimuth. 
The positions of the antennas on the outer ring are slightly rotated, starting at $5^\circ$ azimuth, and placed every $40^\circ$
to accommodate the local terrain.} 
This circular layout has a large number of independent baselines, namely 108, 
to be compared with 120, which is the maximum number of baselines for a 16 element array ($120=16\times15/2$).

We shall consider two surveys with the Tianlai dish array. A mid-latitude survey centred 
at the latitude of the location of the Tianlai array, i.e.  $\delta \sim 44^\circ 10'$,  
composed of a total of 31 constant declination scans, each shifted by 1 degree in declination, spanning the range of 
 $29^\circ 10' < \delta < 59^\circ 10'$, slightly larger than the PAON-4 survey area. 
 At the end of such a survey, Tianlai should be able to provide sky maps over a area of more than 
$5000 \, \mathrm{deg^2}$ of sky and over $\sim 100 \, \mathrm{MHz}$ bandwidth, with a 
sensitivity of $\sim 50 \, \mathrm{mK}$ per $\sim 0.25 \times 0.25 \, \mathrm{deg^2} \times \mathrm{1 MHz}$ pixels.
We also consider a survey of the polar cap area, from the north celestial pole $\delta=90^\circ$ down to $\delta =75^\circ$, in 
16 scans shifted by $1^\circ$ each. The sky area of this survey is about 1/10 of the mid-latitude survey. If completed in the 
same time, the noise level $\sim 3$ times better than the mid-latitude survey $(\sim 17 \, \mathrm{mK})$ can be reached.

\section{Map making for transit interferometers}
\label{sec-transitmapmaking}
In this section we discuss  the specificities of making sky maps from visibilities obtained from a transit type interferometer, 
and the mathematical basis of the method. These issues for transit interferometers 
and the separation of the inversion problem into independent sub-systems using $m$-mode decomposition 
in the spherical harmonic basis have already been discussed in  \citet{2014ApJ...781...57S}. However,
the formalism {\rzrefreq described}  here as well as the corresponding software tools 
\footnote{The code is written in C++ and uses the SOPHYA class library ( http://www.sophya.org ). The GIT repository 
will be available from https://gitlab.in2p3.fr/SCosmoTools/JSkyMap. }
have been developed independently, \rzrefreqa{initially in flat sky approximation, and 
subsequently extended to spherical geometry following \citet{2014ApJ...781...57S}. }

Throughout this section, we shall assume that individual antenna/feed responses, the array geometry and pointing directions
are perfectly known. Moreover we will consider unpolarized sky emission, with brightness or temperature in the
direction $\vec{\hat{n}}$ given by $I(\vec{\hat{n}}) = E^* \, E = | E |^2$ where 
$E(\vec{\hat{n}})$ is the complex scalar emission amplitude in a narrow, nearly monochromatic,
frequency band. 
\rzrefreqa{The method can be extended easily to the case of polarised sky, as shown by \citet{2015PhRvD..91h3514S}. } 
\addarxiv{This is briefly discussed in Appendix \ref{sec-polarisedmapmaking}. }

The Visibility $\mathcal{V}_{ij} \equiv  <s_i^* \times s_j >$ is the \rzrefreqa{short time average} cross correlation of output voltage 
from a pair of antennae or feeds $s_i, s_j$, located at positions $\vec{r_i}, \vec{r_j}$ with $\Delta \vec{r_{ij}} = \vec{r_j} - \vec{r_i}$:
\begin{eqnarray}
s_i & = & \iint E(\vec{\hat{n}}) \, D_i(\vec{\hat{n}}) \, e^{i\vec{k} \cdot \vec{r_i}}  \, d \vec{\hat{n}} \\
\mathcal{V}_{ij} & = & \iint \, I(\vec{\hat{n}}) \, D_i^*(\vec{\hat{n}}) \, D_j(\vec{\hat{n}}) \, 
e^{\vec{k} \cdot  \Delta \vec{r_{ij}} } \, d \vec{\hat{n}} 
\label{eq-visib}
\end{eqnarray} 
where $D_i , D_j$ denotes the complex response function of each feed, 
{ \rzrefreq
$\vec{k}=- \frac{i 2\pi \nu}{c} \vec{n}$ is electromagnetic wave vector at the observation frequency 
$\nu$ and $c$ the speed of light. For arrays with identical feeds pointed to the
same sky direction, $D_i(\vec{\hat{n}})=D_j(\vec{\hat{n}})=D(\vec{\hat{n}})$,  and the visibility expression reduces to } 
\begin{eqnarray}
 \mathcal{V}_{ij} =  \iint \, I(\vec{\hat{n}}) \, L(\vec{\hat{n}}) \,  e^{\vec{k} \cdot  \Delta \vec{r_{ij} } } \, d \vec{\hat{n}}  
\end{eqnarray} 
where $ L (\vec{\hat{n}}) = D^* \, D $ is the antenna primary beam or response in intensity.

\subsection{Classical radio interferometry}

In what we refer here as {\it classical radio interferometry}, {\rzrefreq in the sense that it is familiar to
the majority of radio astronomers,}
a set of identical antennas are used to observe a small region of sky, usually 
to obtain a high resolution image of a source. During the observation period 
all the antennae track the source, compensating the Earth rotation. 
The source intensity $I(\vec{\hat{n}},t)$ and beam response $L (\vec{\hat{n}},t)$ generally varies with time. However, even in the case of constant sources and constant telescope primary beams, the baseline delay $\vec{k} \cdot\Delta \vec{r}(t)$ would still vary with time, due to the rotation of the baseline generated by the 
 rotation of the Earth with respect to the inertial frame of space, as shown in the variation of celestial coordinates of the baseline direction.
  
{\rzrefreq For observations with small field of view, it is possible to use the flat sky approximation in the vicinity 
of the source. 
 For a coplanar array and using the small angle approximation (omitting the so called $w$-term), the visibility is given by
\begin{equation}
  \mathcal{V} (u_0, v_0)  =  \iint I(\xi, \eta) \, L(\xi, \eta) \, 
 e^{ 2 i \pi \left( \xi u_0 + \eta v_0 \right) } \, d \xi ~d \eta 
\label{eq-visib-uv}
\end{equation}
where  $(u_0,v_0)=(\Delta x/\lambda, \Delta y/\lambda)$ is the coordinates of the baseline vector in wavelength units, and $\xi, \eta$ denotes the 
direction cosines of the baseline vector with respect to the reference point. The visibility in this approximation is simply the Fourier transform of 
the sky {\it seen} by a single antenna $I(\xi, \eta) \times L(\xi, \eta)$ for the angular frequencies $(u_0,v_0)$.}
Given the number of available baselines in a real array, and that the baselines of such an array are usually large 
compared to the antenna size, the $\uv$ frequency plane is only sparsely sampled at any moment. 
However, each baseline changes as the antennae follow the source direction 
on the sky, the $\mathrm{(u_0, v_0)}$ follows an arc-shaped track in the $\uv$ plane, enhancing greatly the frequency plane sampling.
It is possible to obtain a local sky map (dirty map) around the targeted position using an inverse Fourier transform. Additional 
processing is required to correct and compensate for the partial coverage of the angular frequencies.
Iterative deconvolution algorithms, e.g. CLEAN  \citep{1974A&AS...15..417H,1980A&A....89..377C}, are applied to recover 
the map of the sky \citep{2007astro.ph..1171S}.
Map of a large area of sky can be obtained by mosaicking of small areas\citep{2007MNRAS.375..625K,2008MNRAS.389.1163M}. However, if the field of view is large,  the $w$-term can not be neglected. A number of formalisms have been 
developed to deal with this, such as faceting \citep{1992A&A...261..353C}, 3D Fourier transform \citep{1999ASPC..180..383P},
 w-projection \citep{2008ISTSP...2..647C}, A-projection\citep{2013A&A...553A.105T}, w-stacking \citep{2014MNRAS.444..606O}, etc. 
{\rzrefreq Other refinement of the CLEAN method have also been developed, such as the the {\it software holography}
\citep{2009MNRAS.400.1814M}  which can deal with direction-dependent beam effects in large field of view interferometer arrays. 
Its application to the analysis of MWA observations can be found in \citep{2012ApJ...759...17S}. }
 

\subsection{Non-tracking transit interferometers}
For interferometers operating in the transit mode, the baselines do not change with time in the ground coordinates, 
at least during an observation period spanning a sidereal day, but the visibilities recorded as a function of 
time correspond to observation of different parts of the sky. We will work in the equatorial coordinates, with right ascension $\alpha$ and 
declination $\delta$. We also introduce the spherical coordinates $(\theta,\varphi)$, with  $\theta = \pi/2 - \delta$ 
and $\varphi=\alpha$.
{\rzrefreq The earth rotation makes the beams time dependent and the effect corresponds to} 
a shift of the beams $L_{ij}(\vec{\hat{n}})$ by an offset angle $\alpha_p(t)$ 
along the right ascension direction:
\begin{eqnarray}
\alpha_p(t) &=& \omega_e \, t \hspace{8mm} t: \mathrm{sidereal \, time} \\
L_{ij}(\vec{\hat{n}},t) & = & L_{ij}((\theta, \varphi),t)  =  L_{ij}(\theta, \varphi - \alpha_p(t))   
\end{eqnarray}
where $\omega_e$ is the Earth angular rotation rate $(2 \pi/ \mathrm{24 \, sidereal \, hours})$.

In the celestial coordinates, the visibility of a baseline at any given time corresponds 
to the convolution of sky with the beam pattern for this baseline $ L_{ij}(\vec{\hat{n}} , t)$. 
Indeed, using discrete time and discrete angular directions on the sky
{\rzrefreq and using $ \left[ \right] $ to denote vectors, }
we can write the vector of visibilities for all baselines and for 
all observation times as a function of the unknown discretized sky $ \left[ I (\vec{\hat{n}} ) \right]  $ and the noise vector:
\begin{eqnarray}
\left[ \mathcal{V}_{ij} (t)\right] & = &  \mathbf{L}_{ij}(t)  \times \left[ I (\vec{\hat{n}} ) \right]   \, + \, \left[ n_{ij}(t) \right]  \label{eq-skylinsys} 
\end{eqnarray}
The beam matrix $\mathbf{L}$ encodes both the array response and the sky scan strategy, 
{\rzrefreq 
$\mathbf{L}_{ij}(t) \sim D_i^*(\vec{\hat{n}} , t) D_j(\vec{\hat{n}} , t)  e^{ i \vec{k} \cdot\Delta \vec{r}_{ij} } $. 
Considering the visibilities for a single narrow frequency band, the $\mathbf{L}$ matrix}
has $N_{\rm pixel}$ columns, 
and $N_t$ (number of time sample) $\times N_b$ (number of baseline) rows. $N_{\rm pixel}$ corresponds to the 
total number of pixels in sky.
If far side lobes can be neglected, one can use a partial map of the sky, limited to the observed region, 
hence decreasing the $N_{\rm pixel}$ and the $\mathbf{L}$ matrix size, 
The determination of the unknown sky $I(\vec{\hat{n}})$ is then the solution of a standard inverse linear problem. 
There are however two difficulties for solving the above equation.
First, the dimension of the matrix $\mathbf{L}$ is very large, typically $ 10^5 \times 10^6 $ for the current generation of experiment, 
and can reach $ 10^6 \times 10^7 $ for the next generation experiments which are being planned, if the intensity 
mapping method proves successful. Indeed, 
the sky brightness unknown vector will have a size of $10^5$ for a resolution 
of a fraction of a degree, determining the number of columns of the $\mathbf{L}$ matrix. CHIME and Tianlai 
will have $\sim 10^3$ baselines and $\gtrsim 10^3$ time samples over 24 hours of observations, leading to $\gtrsim 10^6$ 
rows for the $\mathbf{L}$. Secondly, for many array configurations and sky observation strategies, the linear problem is under-determined 
and a solution can not be unambiguously determined.  

As already shown by \citet{2014ApJ...781...57S}, by working in the space of spherical harmonic 
coefficients and taking advantage of the full circle transit observation strategy foreseen for the intensity mapping 
experiments, the problem can be reduced to a much smaller set of independent  linear systems, one for each spherical $m$-mode.  
The beam pattern associated to each visibility measurement (pair of antenna) is a complex function ( $L_{ij}(\vec{\hat{n}},t) \in \mathbb{C} $ ), and the baseline enters its expression through the phase factor. Its time dependence for transit observations 
is discussed below. Expanding in spherical harmonics and omitting the time dependence of the beam,
\begin{eqnarray}
\label{eq-mapdecomp}
I(\vec{\hat{n}}) & = & \sum_{\ell = 0}^{+\infty}  \sum_{m=-\ell}^{+\ell} \, \mathcal{I}_{\ell , m} \, Y_{\ell,m} (\vec{\hat{n}}) \\ 
L_{ij}(\vec{\hat{n}}) & = & D_i^*(\vec{\hat{n}}) \, D_j(\vec{\hat{n}}) \, e^{i \vec{k} \Delta \vec{r_{ij}} }  \\
& = & \sum_{\ell = 0}^{+\infty}  \sum_{m=-\ell}^{+\ell} \, \mathcal{L}_{ij} (\ell , m) \, Y_{\ell,m} (\vec{\hat{n}})  
\label{eq-beampattern}
\end{eqnarray} 
The spherical harmonics $Y_{\ell,m}$ are defined through the Legendre associated polynomials $P_{\ell}^m(\vec{\hat{n}})$
for which we use the normalisation convention of \citet{Driscoll1994202}
\begin{eqnarray*}
Y_{\ell ,m} ( \vec{\hat{n}})  & = & \sqrt{\frac{ (2 \ell +1) }{ 4 \pi}  \frac{(\ell -m) ! }{(\ell + m) ! } }  \, P_\ell^m ( \cos \theta ) e^{i m \varphi } 
\end{eqnarray*} 

The sky brightness temperature is real, for which the spherical harmonic coefficients satisfy the following 
symmetry relations,
$$ I(\vec{\hat{n}}) \in \mathbb{R} \rightarrow I^* = I \longrightarrow \mathcal{ I} (\ell , -m) = (-1)^m \, \mathcal{I}^* (\ell , m) .$$

{\rzrefreq Given the orthogonality of Spherical Harmonics when integrated over the whole sky, we can express} 
the visibility for a given time $t$ as a sum over the spherical harmonics coefficients. 
{\rzrefreq Expanding both $I(\vec{\hat{n}})$ 
and $ L_{ij}(\vec{\hat{n}}, t)$ in spherical harmonics, use the orthogonality and the above symmetry relation, we obtain}
\begin{eqnarray}
\mathcal{V}_{ij}(t) & = &  \iint \, I(\vec{\hat{n}}) \, L_{ij}(\vec{\hat{n}}, t)  \, d \vec{\hat{n}}  \\
& = &  \sum_{m=-\infty}^{+\infty} \sum_{\ell = |m| }^{+\infty} \,  (-1)^m \, \mathcal{I} (\ell , m) \, \mathcal{L}_{ij} (\ell , -m, t)
\label{eq-vis-alm}
\end{eqnarray}
Notice that we have exchanged the order of the two sums, over $\ell$ and $m$. 
The spherical harmonics coefficients of the rotated/shifted beams can be written as:
\begin{eqnarray}
\mathcal{L}_{ij} (\ell , m, t) & = &  \mathcal{L}_{ij} ^0 (\ell , m)  \, e^{- i m \alpha_p(t)} 
\end{eqnarray}
where  $\mathcal{L}_{ij}^0 (\ell , m)$ denotes the beam spherical harmonics coefficients for the reference $(t=0)$ pointing,
i.e the antenna axis pointing toward $\alpha=0$ right ascension. In the following, we will omit the $^0$ superscript
in the beam coefficients. $\mathcal{L}_{ij}(\ell , m)$ denotes simply the beam for the reference right ascension $\alpha_p=0$.
The recorded visibilities as a function of right ascension $\alpha_p$ can then be expressed as:
\begin{equation}
\mathcal{V}_{ij} (\alpha_p)  =   \sum_{m=-\infty}^{+\infty} \sum_{\ell = |m| }^{+\infty} \,  (-1)^m \, \mathcal{I} (\ell , m) \, \mathcal{L}_{ij}(\ell , -m)  \, e^{i m \alpha_p} 
\end{equation}
We recognise the expression as a Fourier transform for the periodic function $ \mathcal{V}_{ij} (\alpha_p) $; 
{\rzrefreq as the feed response vanishes for large enough $\ell$ ($\mathcal{L}_{ij} (\ell , m) \rightarrow 0$ 
for $\ell > \ell_{\rm max}$), we can write the following relation satisfied by the visibility Fourier coefficients $\tilde{\mathcal{V}}_{ij}(m)$, 
computed from a set a regularly time sampled visibility measurements. } 
\begin{eqnarray}
\tilde{\mathcal{V}}_{ij}(m)  =  \sum_{\ell = |m| }^{+\ell_{\rm max} } \,  \,  (-1)^m \, \mathcal{I} (\ell , m) \mathcal{L}_{ij} (\ell , -m) 
\label{eq-vis-I-FFT}
\end{eqnarray}
The $m$-mode of the visibility  for both  positive and negative m  ($\pm m$)  is given by
sky spherical harmonics coefficients of the same $m$,
\begin{eqnarray} 
\tilde{\mathcal{V}}_{ij}(m) & = & \sum_{\ell = |m| }^{+\ell_{\rm max} } \,  \,  (-1)^m \, \mathcal{ I} (\ell , m) \mathcal{ L}_{ij} (\ell , -m)  \label{eq-mmode1} \\
\tilde{\mathcal{V}}^*_{ij}(-m) & = & \sum_{\ell = |m| }^{+\ell_{\rm max} } \,  \,   \mathcal{ I} (\ell , m) \mathcal{ L}^*_{ij} (\ell , m) \label{eq-mmode2}
\end{eqnarray} 
{\rzrefreq The full linear system of Eq.~(\ref{eq-skylinsys}) can thus be decomposed into a set of 
much smaller ($10^3 \times 10^3$) independent linear system, one for each $m$, with $m_{\rm max} = \ell_{\rm max}$. 
The beam matrix $\mathbf{L}$ has indeed a block diagonal structure in the harmonic space. 
Grouping all array baselines together in a vector, and taking into account the noise contribution,
the visibility measurement equation in the Fourier space can be written in matrix form as: }
\begin{eqnarray}
\left[  \tilde{ \mathcal{V} }  \right]_m & = &   \mathbf{L}_{m} \,  \times \, \left[  \mathcal{I} (\ell) \right]_m +  \left[  \tilde{n}  \right]_m
\label{eq-visibility}
\end{eqnarray}
{\rzrefreq The sky spherical harmonics coefficient for a given $m$ and for $m \leq \ell \leq \ell_{\rm max}$ are grouped 
in the sky vector $\left[  \mathcal{I} (\ell) \right]_m$. 
We will consider only positive $m$ values ($0 \leq m \leq \ell_{\rm max}$) for the linear systems defined above,  
the two visibility measurements for $\pm m$ of equations \ref{eq-mmode1} and \ref{eq-mmode2} 
will be represented by two rows of the matrix $\mathbf{L}_m$. This matrix will thus have 
$\ell_{\rm max}$ columns and $2 \times n_{\rm beams}$ rows.  
The total number of beams $n_{\rm beams}$ will be more precisely defined in the next paragraph. The $  \left[  \tilde{n}  \right]_m $ 
represent the noise contribution vector to the $m$-mode  visibilities, corresponding to the Fourier transform of time domain noise. }

 
For dish arrays, the instantaneous field of view is a small fraction of the whole sky, and a circular strip of sky along one of the 
latitude line can be obtained by carrying out transit observation for 24 sidereal hours continuously. 
By changing the elevation angle of the dish pointing, strips with different central declination can be obtained.  
For dish arrays, the effective number of beams would be equal to the number of different baselines times the number of constant elevation scans, 
$$n_{\rm beams}  =  N_b \times n_{\delta_p} . $$
The beam for an  antennae pair $ij$ making constant elevation drift scan observation with declination $\delta_p$ is
\begin{eqnarray} 
L_{ij}^{\delta_p} & = & D_i^{\delta_p}(\hat{\vec{\hat{n}}}) D_j^{\delta_p *}(\vec{\hat{n}}) e^{i\vec{k}\cdot \vec{\Delta r_{ij}}}  \\
& = & \sum_{lm} \mathcal{L}^{\delta_p}_{ij}(\ell,m) Y_{\ell m}(\vec{\hat{n}})
\end{eqnarray}

\subsection{Solving the system}
{\rzrefreq The sky brightness temperature spherical harmonics coefficients can be estimated by solving each of 
the $m$-modes linear systems defined by Eq. \ref{eq-visibility} . 
The $\mathbf{L}_m$ matrix size is $2 n_\mathrm{beams} \times \ell_{\rm max}$, with 
$\ell_{\rm max}$  around few thousands for array sizes $\lesssim 100 \, \mathrm{m}$ and 
a number of beams up to a to a few thousands for the current generation of instruments. 
Although these systems are usually under-determined, the solution can formally be written as: } 
\begin{eqnarray}
\left[ \widehat{\mathcal{I}}(\ell)  \right]_m &=&  {\large \mathbf{H}_m } \, \left[ \tilde{\mathcal{V}}  \right]_m  
\end{eqnarray}
{\rzrefreq where $\left[ \right]$ are used to denote vectors and $\mathbf{H}_m$ is the noise weighted Moore-Penrose pseudo-inverse of $\mathbf{L}_m$\citep{2012BrJPh..42..146B}. }

To make map from a given set of visibilities with noise, we look for a maximum likelihood solution. Here we 
assume that the noise on visibility measurement follows a Gaussian random process, with variance
$\mathbf{N}_m=< \left[ \tilde{n} \right]_m \, \left[ \tilde{n} \right]_m ^\dagger>$. {\rzrefreq 
We consider moreover that noise is uncorrelated for different $m$-modes. This hypothesis 
is valid as long as the time domain noise is a Gaussian random process characterised by a power spectrum.} 
The solution is given by
\begin{eqnarray}
\mathcal{\widehat{I}}_m  &=& (\mathbf{L}_m^\dagger \mathbf{N}_m^{-1} \mathbf{L}_m)^{-1} \mathbf{L}_m^\dagger \mathbf{N}_m^{-1} \mathcal{V}_m  
\equiv \mathbf{H}_m \,  \mathcal{V}_m  \\
\label{eq-inverse}
\mathbf{H}_m &=& (\mathbf{L}_m^\dagger \mathbf{N}_m^{-1} \mathbf{L}_m)^{-1} \mathbf{L}_m^\dagger \mathbf{N}_m^{-1} 
\label{eq-Bm-full}
\end{eqnarray}
Redundant baselines are counted once with their noise level 
being scaled accordingly, i.e. $\sigma_n^2 \propto N_{rb}^{-1}$, where $N_{rb}$ denotes the number of redundant baselines
(number of antennae pairs with the same baseline). 

If we further assume that  noise is uncorrelated between different baselines, the noise covariance matrix $\mathbf{N}_m$ 
for each $m$ becomes diagonal.
In this case, the computation can be further simplified as
\begin{eqnarray}
\mathbf{H}_m &=&   \left( \mathbf{N}_m^{-\frac{1}{2}} \, \mathbf{L}_m \right)^{-1} \,  \mathbf{N}_m^{-\frac{1}{2}}
\label{eq-Bm}
\end{eqnarray}
\rzrefreqa{ The pseudo-inverse is computed using the Singular Value Decomposition (SVD). An absolute threshold, and 
a second relative threshold defined as a fraction of the largest eigenvalue are used to avoid numerical instabilities during inversion. 
Eigenvalues below the threshold, as well as their inverse are simply put to zero.  
A review on the Moore-Penrose pseudo inverse computation and properties can be found in \citet{2012BrJPh..42..146B}. 
Once all the sky spherical harmonics coefficients are determined by solving all the $m$-modes systems, }
we can compute the sky map $\widehat{I}(\vec{\hat{n}})$ by performing an inverse Spherical Harmonics Transform (SHT)
on the estimated spherical modes coefficients $[\widehat{\mathcal{I}}(\ell)]_m$.

%

\subsection{Reconstructed maps and PSF}
\label{sec-synbeam}
The $\mathbf{L}_m$ matrices depend only on the array configurations or baselines, individual antenna beams 
and the scanning strategy, so $\mathbf{H}_m$ also depends on these. It does depend on the noise covariance matrix structure, 
but not on its values. For instance, the $\mathbf{H}_m$ remains unchanged if we change the total survey duration, or the system temperature for all feeds, without changing the array 
configuration (baselines and number of redundant baselines), or the scanning strategy, i.e. how the fraction of the total survey 
time spend on each declination. So, once the $\mathbf{H}_m$ are computed, we can apply them to different input visibilities 
to reconstruct different sky maps:
\begin{itemize}
\item To obtain the instrument {\rzrefreq response to a point source, which corresponds to the PSF (Point Spread Function) or the instrument beam, 
we reconstruct the maps from mock visibilities computed from input maps containing point sources at different declinations. 
The PSF is independent of the right ascension, but varies for different declinations.}
\item Starting from an input sky map, we can compute its decomposition into spherical harmonics 
($\mathcal{I}(\ell,m)$) using the SHT.  Then using the $\mathbf{L}_m$ matrices computed by the map-making tools,
we can compute the visibility matrices, with or without adding noise. Applying the $\mathbf{H}_m$ to a set of such mock 
visibility data, we can reconstruct the sky maps as seen by an transit interferometric array. For demonstrations in this paper,
we have used the Leiden-Argentina-Bonn(LAB) survey \citep{2005A&A...440..775K} for the sky emission at 
21--cm (Galactic HI). 
The LAB data has been used to create spherical maps at several frequencies, suitable for processing by our software tools. 
For the radio continuum which is dominated by the Galactic synchrotron emission at the relevant frequencies,  
we have used full sky maps generated by the Global Sky Model (GSM) \citep{2008MNRAS.388..247D},
\item We can also compute pure noise maps, if the input visibility vectors contain contribution from noise only. 
These pure noise maps can be used to compute survey noise power spectrum, as an alternative to using the noise 
covariance matrix (see section \ref{sec-errorcovar} below). To limit statistical fluctuations, we generated 50 random 
noise maps from noise-only visibilities for computing noise power spectra. 
\end{itemize}

We have used spherical maps with HEALPix pixelization scheme \citep{2005ApJ...622..759G} for the 
reconstructed maps presented in this paper, although two other pixelization schemes are currently provided 
by SOPHYA and could be used by the map making software. 
We have checked that the results are not sensitive to $\ell_{\rm max}$ and the corresponding HEALPix $\mathrm{n_{\rm side}}$ 
parameter as long as the map resolution is at least a factor 2 higher than the synthesised  beam resolution:
$$ \ell_{\rm max} \gtrsim \frac{2 \pi \, D_{\rm array}}{\lambda} $$ 
where $D_{\rm array}$ is the diameter of the disk covering the full array.
For PAON-4 with $D_{\rm array} \sim 18 \, \mathrm{m}$, $\ell_{\rm max}=750$ and $\mathrm{n_{\rm side}}=256$ would be more than 
enough for reconstructing maps. However, The Tianlai circular array configuration with $D_{\rm array} \sim 40 \, \mathrm{m}$ 
requires $\ell_{\rm max} \gtrsim 1200$. We have thus used $\ell_{\rm max}=1500$ and HEALPix $\mathrm{n_{\rm side}}=512$, 
corresponding to a pixel resolution of $\sim  6.9 \, \mathrm{arcmin}$, for most of the results presented in this paper.

\subsection{Instrument response and transfer function}
\label{sec-transferfunc}

The $m$-mode reconstruction matrix  $\mathbf{R}_m \equiv(\mathbf{H}_m \mathbf{L}_m)$ tells us how the estimated 
sky spherical harmonics coefficients $(\widehat{\mathcal{I}}(\ell,m) )$ are related to the true sky ones 
$(\mathcal{I}(\ell, m) )$ 
\begin{eqnarray}
\left[ \widehat{\mathcal{I}}(\ell)  \right]_m & = & \mathbf{R}_m \, \left[ \mathcal{I}(\ell)  \right]_m
\end{eqnarray}
Ideally, if $\mathbf{R}_m = \mathbf{I}$ where  $\mathbf{I}$ is the identity matrix, then we would be able to recover the
sky spherical harmonic $m$-mode completely from the observations. However, in reality this is not possible. 
Although each $m$ mode is measured  independently for a full circle transit observation, for each given $m$
the different $\ell$ coefficients are still correlated, the physical measurement data is a mix of 
different $\ell$ mode contributions. The $\mathbf{R}_m$ matrix gives the
window function in $\ell$-space for the estimated sky. We can define the \rzrefreqa{core} response matrix $\mathbf{R}$  
by extracting the diagonal terms from individual $\mathbf{R}_m$ matrices:
\begin{eqnarray*}
\mathbf{R}(\ell, m) & = & \mathbf{R}_m (\ell , \ell) 
\label{eq-Rmatrix} 
\end{eqnarray*}
For reconstruction,  the $\mathbf{R}(\ell,m)$ is insufficient and the original $\mathbf{R}_m$ matrices are needed,
but the $\mathbf{R}(\ell,m)$ matrix can give some idea of how well an $(\ell, m)$ mode is measured with the given 
array, so it can help us to see the effectiveness of our reconstruction in the $(\ell,m)$ space. 

We can further compress the response function by computing the transfer function, which is defined by the 
average over the $m$-modes from the response matrix $\mathbf{R}$,
\begin{eqnarray}
T(\ell) & = & \langle | \mathbf{R}(\ell, m) | \rangle_m 
\end{eqnarray}
Let's consider visibilities corresponding to an input white noise map, without any additional noise ($\sigma_{\rm noise}=0$)
\begin{eqnarray*}
\langle | \mathcal{I}(\ell,m) |^2  \rangle  & = & C^\mathrm{in}(\ell) = \mathrm{const}  \\
 \langle   \mathcal{I}(\ell,m)  \, \left( \mathcal{I}(\ell',m') \right)^* \rangle & = & \delta_{\ell \ell' , m m'} \,   C^\mathrm{in}(\ell)  
\end{eqnarray*} 
if we reconstruct the map from such visibilities and computed the reconstructed map power spectrum,
we can write it as:
\begin{eqnarray*}
\langle \left[ \widehat{\mathcal{I}}(\ell) \right]_m \, \left[ \widehat{\mathcal{I}}(\ell') \right]_m^\dag \rangle & = & 
\langle \mathbf{R}_m \left[ \mathcal{I}(\ell)^{in} \right]_m \left[ \mathcal{I'}(\ell)^{in} \right]_m^\dag \, \mathbf{R}_m^\dag \rangle 
\end{eqnarray*} 
where $\dag$ denotes Hermitian conjugate (transpose and complex conjugate). Noting that $\mathbf{R}_m$ are 
projector matrices \needcitation. 
\begin{eqnarray*}
\mathbf{R}_m^\dag & =& \mathbf{R}_m \\
\mathbf{R}_m^\dag \, \mathbf{R}_m & = & \mathbf{L}_m^\dag \mathbf{H}_m^\dag \,  \mathbf{H}_m \mathbf{L}_m 
=  \mathbf{H}_m \mathbf{L}_m = \mathbf{R}_m 
\end{eqnarray*}
and that for a white noise input map, the covariance matrix in spherical harmonics space is proportional to the 
identity matrix $\mathbf{I}$:
$$ \langle \left[ \mathcal{I}(\ell) \right]_m \left[ \mathcal{I}(\ell') \right]_m^\dag \rangle = \mathrm{const}   \times \mathbf{I} $$ 
we obtain that:
$$ \langle | \widehat{\mathcal{I}}(\ell,m) |^2  \rangle  = \mathbf{R}_m(\ell, \ell) \, C^\mathrm{in}(\ell)  = \mathbf{R}(\ell, m) \, C^\mathrm{in}(\ell) $$
So, if we compute the reconstructed map power spectrum by averaging $ | \widehat{\mathcal{I}}(\ell,m) |^2 $ over all m-modes,
the ratio of the reconstructed map angular power spectrum, to the input map, flat angular power spectrum would be equal
to the transfer function defined above:
$$ C^{\mathrm{rec}}(\ell)  =  \langle | \widehat{\mathcal{I}}(\ell,m) |^2 \rangle_m  \longrightarrow
T(\ell)  =   \frac{C^{\mathrm{rec}}(\ell)}{ C^\mathrm{in}(\ell)  } $$
where $C^\mathrm{rec}(\ell)$ is the power spectrum of the reconstructed map, computed from visibilities 
corresponding to the observation of a white noise sky, and  $C^\mathrm{in}(\ell) = \mathrm{const} $ denotes
the input sky flat power spectrum.
The computation of the transfer function from reconstructed map power spectrum 
proves easier to use when additional filtering in the $(\ell,m)$ plane or masking in angular space is applied
after the $\hat{\mathcal{I} }(\ell, m)$ computation stage. 

{\rzrefreq 
If we consider a masked sky map, a spherical map where pixels outside the observed area are put to zero, 
the computed variance of pixel values is lowered by a factor $\sim f_{\rm sky}$, where $ f_{\rm sky} = \Omega_\mathrm{obs}/(4 \pi)$ 
is the observed fraction of the full sky area. We expect thus to obtain transfer functions with levels close 
to $f_{\rm sky}$.  

It should be noted that the cosmological signal is characterised by its 3D power spectrum $P(k)$, which 
determines the signal power spectrum $C^\mathrm{sig}(\ell)$ for frequency shells reconstructed by the map making process. 
The transfer function can be used to compute the {\it observed signal} power spectrum for each frequency shell, 
$C^\mathrm{obs}(\ell)=T(\ell) \times C^\mathrm{sig}(\ell)$. In the absence of foreground, the comparison between the 
expected observed signal power spectrum $C^\mathrm{obs}(\ell)$ and the noise power spectrum  
(section \ref{sec-errorcovar} below) is the main tool to estimate the ability of a given instrument to measure 
a signal characterised by its power spectrum. 
}


\subsection{Error covariance matrix and noise power spectrum}
\label{sec-errorcovar}

If we consider the reconstruction of sky spherical harmonics coefficients from pure noise visibilities
$(\tilde{\mathcal{V}}_{ij} = \tilde{n}_{ij} )$, the covariance 
matrix $\mathbf{Cov}_m (\ell_1 , \ell_2)$ of the estimator $\widehat{\mathcal{I}}(\ell,m)$ for each mode $m$ 
can be computed from the $\mathbf{H}_m$ matrix and the noise covariance matrix:
\begin{eqnarray*}
\mathbf{N}_m & = & \left[ \tilde{\mathcal{V}}_{ij} \right]_m \cdot \left[ \tilde{\mathcal{V}}_{ij} \right]_m^\dag  \\
\mathbf{Cov}_m (\ell_1 , \ell_2) & = & \langle \left[ \widehat{\mathcal{I}}(\ell) \right]_m \cdot \left[ \widehat{\mathcal{I}}(\ell) \right]_m ^\dag \rangle\\
& = & \mathbf{H}_m \, \mathbf{N}_m \,  \mathbf{H}_m^\dag 
\end{eqnarray*}

The covariance matrix is not diagonal, especially due to partial sky coverage in declination. 
However, if we ignore this correlation and use the diagonal terms only for each $m$ mode, we can gather them 
together to create the $ \sigma_{\mathcal{I}}^2(\ell,m)$ variance matrix. This matrix informs us on how well each 
$(\ell,m)$ mode is measured. This noise variance matrix can then be used in subsequent processing steps,
for example to throw out modes with large errors, or by applying weights inversely proportional to the error variance.
We can even compress further this information by computing the noise power spectrum:
\begin{eqnarray}
\sigma_{\mathcal{I}}^2(\ell,m)  & = & \mathbf{Cov}_m (\ell , \ell)  \label{eq-covardiag} \\
C^{\mathrm{noise}} ( \ell ) & = & \langle \sigma_{\mathcal{I}}^2(\ell,m)  \rangle_m   \label{eq-noise-cl} 
\end{eqnarray}
As pointed out in the previous section, we can also compute noise maps by applying the $\mathbf{H}_m$ matrices to noise 
only visibility. We can then use the reconstructed maps to compute the noise power spectrum, which is identical to the one 
obtained directly from Eq.(\ref{eq-noise-cl}). However, the noise maps can still prove useful for computing the 
noise power spectrum when further filtering in angular or spherical harmonics space is applied.

As mentioned earlier, the $\mathbf{H}_m$ matrix does not change if the visibility noise matrix is scaled.
The map making is thus performed with a value of visibility time sample noise $\sigma_{\rm noise}=1 \mathrm{K}$. 
To compute the noise level for a given survey, the noise covariance matrix is rescaled by the effective $\sigma_{\rm noise}$.
To compute this value, we take into account the system temperature $T_{\rm sys}$, the total survey time $t_{\rm survey}$
and the frequency band $\Delta \nu$ for each sky map. The number $n_t$ of time samples for the visibilities 
for a 24 hours constant declination scan is fixed by the maximum value of $m$, with $n_t=2 m_{\rm max}$,
as the $m$-modes visibility vector time indexed visibilities are related by an FFT. $m_{\rm max}$ is itself 
equal to the $\ell_{\rm max}$ which has to be chosen so that $\ell_{\rm max} \gtrsim \frac{2 \pi D_{\rm array}}{\lambda}$. 
For the survey strategies discussed in this paper, we have distributed the observation time evenly among 
all constant declination scans. The  effective $\sigma_{\rm noise}$ for measured visibility time samples can then be 
written as a function of integration time per time sample $t_{\rm int}$:
\begin{eqnarray}
\sigma_{\rm noise}^2 & = & \frac{2 T^2_{\rm sys}}{t_{\rm int} \, \Delta \nu}   \label{eq-signoise-1}  \\
t_{\rm int} & = & \frac{ t_{\rm survey} / \mathrm{Days} }{n_{\delta_p}}  \times \frac{24 \times 3600}{2 \, m_{\rm max} }  \label{eq-signoise-2} 
\end{eqnarray} 

{\rzrefreq 
As pointed out above, the variance of masked map is lower by a factor $f_{\rm sky}$ compared to the corresponding 
full map. In order to make the noise power spectra comparable for the different configurations, all the noise power 
spectra shown in this paper are rescaled according to $ C^{\mathrm{noise}} ( \ell ) \times (1/f_{\rm sky}) $, where  
$ C^{\mathrm{noise}} ( \ell )$ is computed from eq. \ref{eq-noise-cl}.
}

%

\subsection{Filtering in $(\ell, m)$ space and angular masking}
\label{sec-lm-filtering}
Once the sky spherical harmonic coefficients are computed by the map making process described above, 
it is possible to apply additional filters, either in the Fourier space ($(\ell, m)$ plane) or angular space. 
These filters can be used for example to decrease the noise level in the final sky map, by ignoring or damping 
modes with high noise, the noise variance matrix being the key tool to design such filters.
Another possible application would be to correct the instrument response frequency dependence, in which case 
the $\mathbf{R}$ matrix or the full response matrix $\mathbf{R}_m$ would be the key tool.

Finally, optimal filters for component separation and cosmological signal extraction could be designed 
by the simultaneous use of the instrument response, noise covariance matrix, statistical knowledge 
of the signal and the foregrounds. The discussion of such optimal filtering methods is beyond the scope 
of this paper. Here, we will apply a simple mask with sharp edges in declination to define precisely the fiducial sky region, 
and we also consider two simple noise-reduction filters described below.

The first simple filter  we consider is $W_1(\ell,m)$,  which uses the noise variance matrix $\sigma^2_\mathcal{I}$. 
\begin{equation*}
W_1(\ell,m)  = \left\{\begin{array}{lll}
1, & \hspace{2mm} \mathrm{if}  \hspace{2mm}  & \sigma^2_\mathcal{I}(\ell, m) < \sigma^2_{\rm thr} ; \\
\frac{1}{\sigma^2_\mathcal{I}(\ell, m)}, & \hspace{2mm} \mathrm{if}  \hspace{2mm} & 
\sigma^2_\mathcal{I}(\ell, m) > \sigma^2_{\rm thr}.\end{array}\right.
\end{equation*} 
The threshold $\sigma^2_{\rm thr}$ is defined as $ K \,   \sigma^2_{\rm min} $ where $ \sigma^2_{\rm min} $ denotes 
the minimum value of the noise variance matrix, and $K$ a constant factor. A value of $K = 50$ has been 
used for the examples shown in this paper.  This filter suppresses modes with very large errors. 

A second filter we consider is the weight function $W_2(\ell)$, independent of $m$. This second 
weight function is used to reject high noise modes at high $\ell$ near the edge of instrument sensitivity region,
and also the low $\ell$ modes when the autocorrelation signal is not used: 
\begin{equation*}
W_2(\ell)  = \left\{ \begin{array}{ll}
 \left( 1 + e^{  \frac{\ell - \ell_A}{\Delta \ell_A}  } \right)^{-1}, &\mathrm{With-AutoCorr} ; \\
 \left( 1 + e^{  \frac{\ell - \ell_A}{\Delta \ell_A}  } \right)^{-1}  \times \left( 1 + e^{  \frac{\ell_B - \ell}{\Delta \ell_B}  } \right)^{-1}  , 
 &\mathrm{No-AutoCorr}. \end{array}\right.
\end{equation*} 
{\rzrefreq The filter parameter $\ell_A, \Delta \ell_A, \ell_B, \Delta \ell_B$ are determined empirically}.  
For the case of PAON-4,  
$\ell_A = 440 \, , \, \Delta \ell_A=15$,  $\ell_B = 90 \, , \, \Delta \ell_B=10$.
For the Tianlai circular dish array case ($\ell_A = 1050 \, , \, \Delta \ell_A=15$,  $\ell_B = 120 \, , \, \Delta \ell_B=10$) .



\section{Application to PAON-4}
\label{sec-paon4case}

\begin{figure*}
\centering
\includegraphics [width=0.8\textwidth] {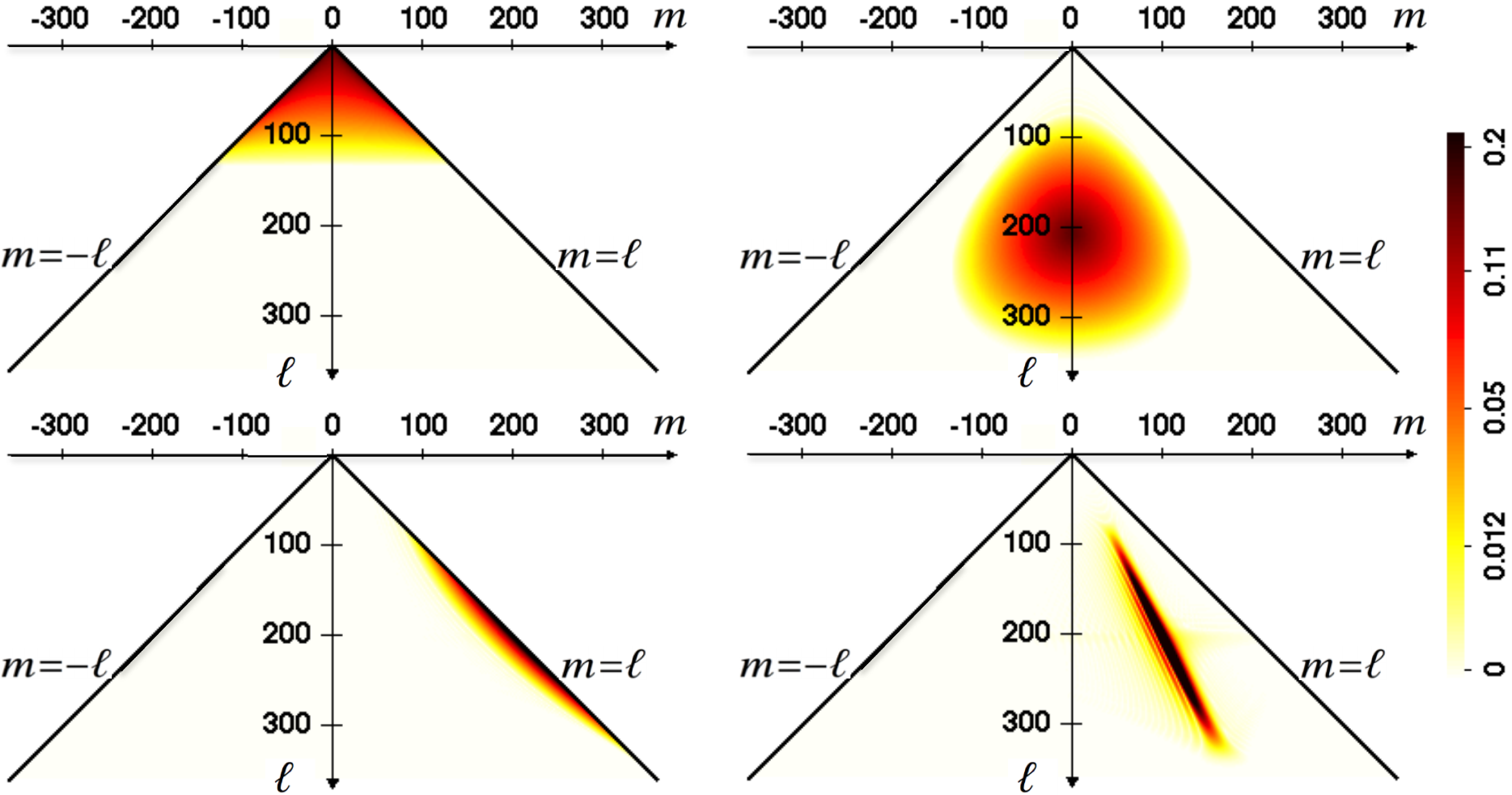}
\caption{The beam patterns in spherical harmonics $\mathcal{L}_{\ell,m}$ with dish size $D=4.5m$. Top left: auto-correlation for a dish pointing toward equator $\delta=0^\circ$; top right: cross-correlation beam for an NS baseline with $d_{\rm sep}=7$m at $\delta=0^\circ$; bottom left: cross-correlation beam for an EW baseline with $d_{\rm sep}=7$m at $\delta=0^\circ$; bottom right: cross-correlation beam for an EW 
baseline with $d_{\rm sep}=7$m at $\delta=60^\circ$. }
\label{fig-beams}
\end{figure*}

\begin{figure}
\centering
\includegraphics[width=0.5\textwidth]{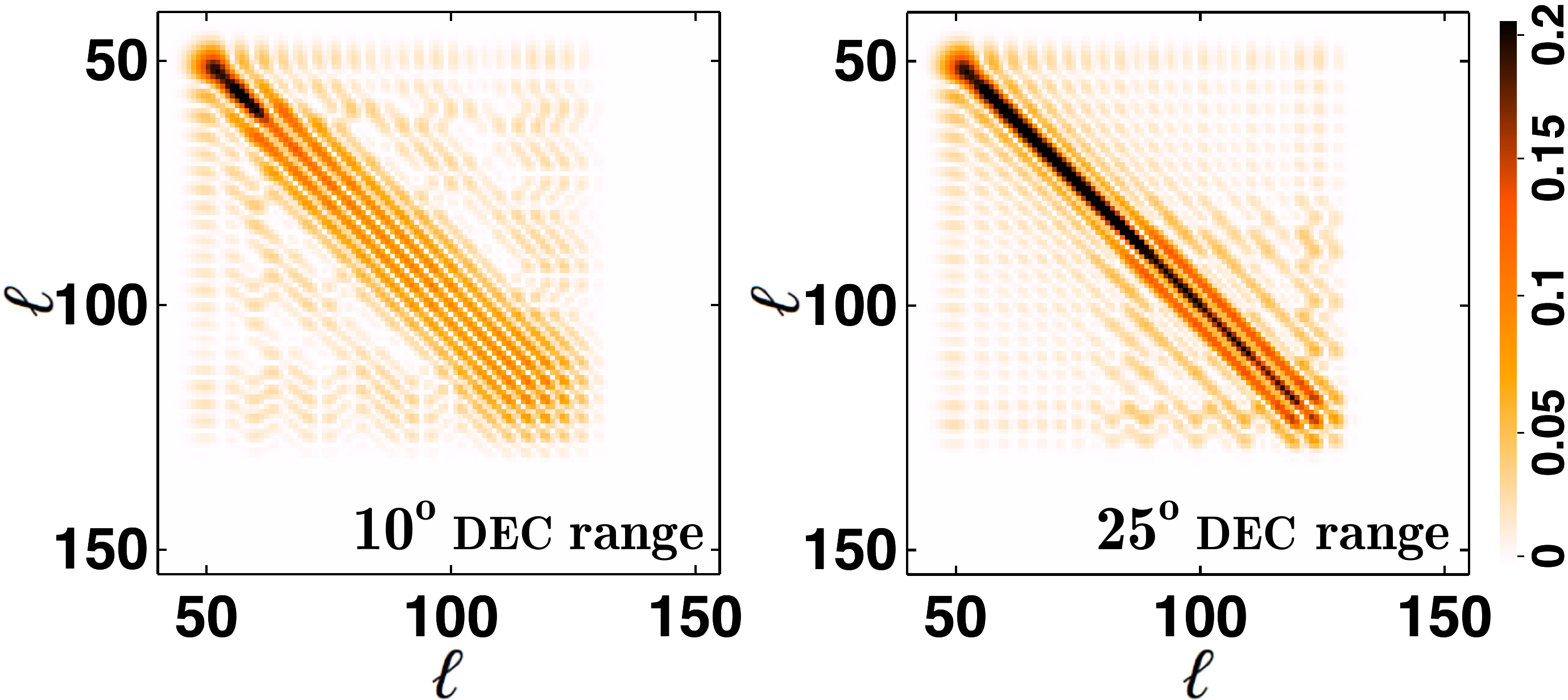}
\caption{Examples of $\mathbf{R}_m$ matrix for a single dish ($D_\mathrm{eff}=4.5 \, \mathrm{m}$) and $m=40$: survey of 
$\sim 10^\circ$ declination band (left) and $\sim 25^\circ$ (right), centred around $\delta = 40^\circ$  } 
\label{fig-Rm_matrix}
\end{figure}

In this section,  we apply the formalism developed in the last section to the PAON-4 case, and also compare it with the 
compact $2\times 2$ array and a large single dish.  In order to provide clues for understanding a given instrument response 
and the impact of various parameters, 
we first discuss the features of response matrices $\mathbf{R}_m , \mathbf{R}$ in few specific cases, 
in particular  a non-compact $2\times 2$ array. This will also illustrate why we choose the very 
compact array layout for PAON-4 and Tianlai. 

\subsection{Beams and response matrix features}
\label{sec-paon4respmtx}

The antenna pair beam patterns are the key elements to understand the complete instrument response. 
Figure \ref{fig-beams} shows the beam patterns $\mathcal{L}(\ell,m)$ for a few configurations (baselines and declinations)
for dishes with effective {\rzrefreq diameter} $D_\mathrm{eff}=4.5 \, \mathrm{m}$ and $\lambda=21\, \mathrm{cm}$. 
The top-left panel of Fig. \ref{fig-beams} shows the single dish auto-correlation
with the antenna axis pointed due south to the equator ($\delta = 0$). 
We see that the beam pattern covers a triangular shaped area in the $(\ell,m)$ space centred at $m=0$. 
Due to pointing at the equator, the beam coverage extends to the maximum allowed m value, i.e. $m= \ell $. 
For pointing to an arbitrary direction defined by the declination $\delta$, the bound would actually be $m= \ell \cos\delta$.
As expected, we can see also that the beam falls off beyond $\ell_{\rm max} \sim 2 \pi D_\mathrm{eff} / \lambda \sim 135$.

For cross correlations, we expect the beam in $(\ell,m)$ space to be centred at 
$(\ell,m)=(2\pi |\vec{u}|, 2\pi u \cos\delta)$, where $\vec{u}\equiv \mathrm{(u,v,w)}$ is the baseline vector in wavelength units. 
For an antenna pair separated by an east-west (EW) baseline, the beam has a crescent shape, with $m \approx \ell \cos\delta$ 
and centred at $(\ell_0 \sim 2 \pi u, m \sim \ell_0 \cos \delta)$. This is shown 
on the bottom left and bottom right panels, for an east-west baseline  with length $d_\mathrm{sep}=7$m and 
for declinations $\delta=0^\circ$ and $\delta=60^\circ$ respectively. 
By contrast, the north-south (NS) baseline  (top right panel) is only mildly 
sensitive to the sky intensity variations along the EW direction; the beam pattern is centred at $m=0$ and 
$\ell_0 \sim 2 \pi d_\mathrm{sep} / \lambda \sim 210$,
the extension along m-direction $(- m_{\rm max} < m < + m_{\rm max})$ is given by the dish size
$m_{max} \sim 2 \pi D_\mathrm{eff} / \lambda \sim 135$.

{\rzrefreq 
To gain a better sense of the reconstruction, we plot in Fig.\ref{fig-Rm_matrix} the $\mathbf{R}_m$ matrix for a single dish with 
effective diameter $D_\mathrm{eff}=4.5 \, \mathrm{m}$ observing in transit mode. We have shown two matrices for $m=40$,
the left panel corresponding to a survey of $\sim 10^\circ$ declination band in the range $35^\circ \lesssim \delta \lesssim 45^\circ$, 
while the right panel shows the $\mathbf{R}_m$ for a wider $\sim 25^\circ$ survey in the range $35^\circ \lesssim \delta \lesssim 60^\circ$. 
We can see that the $\ell$ response starts at a minimum $\ell$-value corresponding to $m / \cos \delta_{\rm min}$, equal to 
$\ell \sim 49$ for $m=40$ and $\delta_{\rm min}=35^\circ$ and extends up to $\ell \sim 135$, determined by the dish size. 
The width of the diagonal band which determines the $\ell$ resolution, starts by decreasing, going through a minimum 
around $\ell = m / cos \delta_{rm max}$ ($\ell \sim 80$ here) and then increases reaching its maximum near the end of $\ell$ sensitivity range. 
As expected, one can see that the $\ell$ resolution gets enhanced by a wider declination coverage, when comparing the 
$\mathbf{R}_m$ matrix shown on the right panel ($\Delta \delta \sim 25^\circ$) with the left panel ($\Delta \delta \sim 10^\circ$). }
The response matrix becomes diagonal for a survey covering the full sky.  

\begin{figure}
\centering
\includegraphics [width=0.4\textwidth] {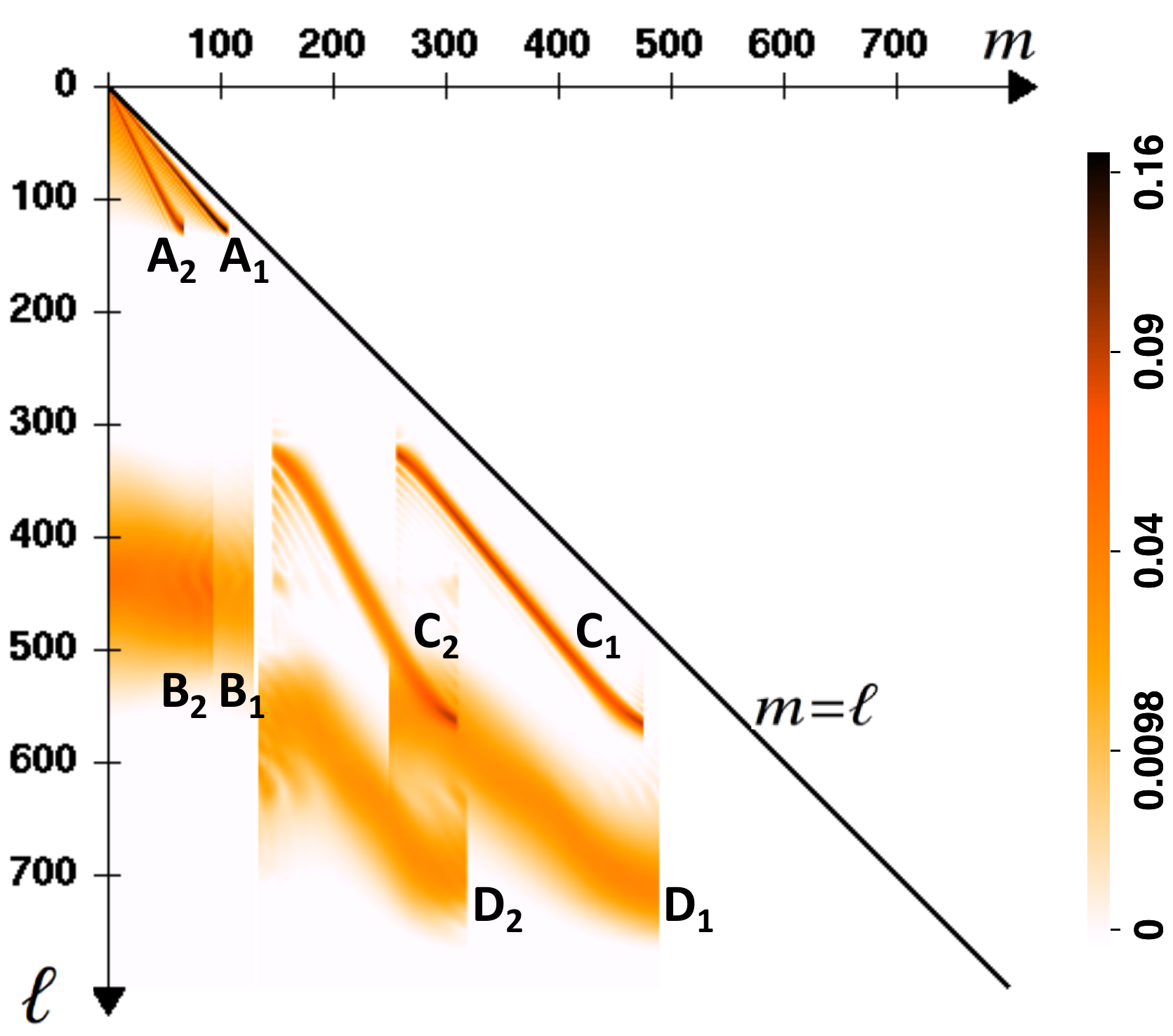}
\caption{The $\mathbf{R}$ matrix for 4 dishes with 15-m baselines and pointing at $\delta_1=35^\circ 23'$ and $\delta_2=59^\circ 23'$. }
\label{fig-2strip}
\end{figure}

Below, we shall analyse the \rzrefreqa{core} response matrix $\mathbf{R}$ which gathers the diagonal terms of $\mathbf{R}_m$
(Eq. \ref{eq-Rmatrix}). 
As an example, let us consider a $2\times 2$ array here, where four dishes of 5 m {\rzrefreq diameter} 
similar to those used in PAON-4  
are arranged on the four corners of a square, with the side length of the square to be 15 m.
The visibility of the two diagonal baselines are related by complex conjugate in the $(\ell, m)$ space: 
$ V_{\rm SE-NW}(\ell,m) = V_{\rm SW-NE}(\ell,-m)=V_{\rm SW-NE}^\dagger(\ell,m)$, 
so in Fig.\ref{fig-2strip} we will see these two baselines appear to occupy the same region in the $(\ell,m)$ space.

To show how $\mathbf{R}(\ell,m)$ matrix will look like for the transit observation of a narrow strip along a constant 
declination line, as would be achieved by a single pointing of the dish array, and show also the effect of observations 
at  different declinations, we plot the $\mathbf{R}$ matrix in logarithmic colour scale in Fig.~\ref{fig-2strip},  
for two constant declination scans, one at $\delta_1=35^\circ 23'$ and another at $\delta_2=59^\circ 23'$.
These two declinations correspond to the edges of the sky region that would be covered by the PAON-4 observations. 
We can easily see on the figure two sets of covered regions, corresponding to the two declinations.  
Each baseline for each pointing covers one distinct region in the $(\ell,m)$ space, as expected from the beam 
shapes discussed above (fig. \ref{fig-beams}). 
We can distinguish four pieces  in Fig. \ref{fig-2strip}:
the wing-shaped \rzrefreqa{A} region near the origin $(\ell, m)=(0,0) $ are derived from the auto-correlation, with the two intensive
stripes of $m=\ell \cos\delta$ for the two declinations, the $35^\circ 23'$ one  \rzrefreqa{(marked as $A_1$) on the {\em outer} side. }
The \rzrefreqa{B} region around $(\ell = 450, m=0)$ obviously corresponds with the NS baseline with $d_{\rm sep}=15$m. 
Here the two pointing directions are largely coincident with each other, except that the $35^\circ 23'$ one ($B_1$) extends further in the $m$ direction. 
We also see that the region spread is $\Delta \ell \sim 150$, so $\Delta \ell/\ell_{c} \sim D/d_{\rm sep}$. 
The EW baseline corresponds to the two narrow strips \rzrefreqa{C}  centred at the same $\ell$ but with $m=\ell \cos\delta$, with some fringes 
within the region. 
The diagonal baselines correspond to the \rzrefreqa{D} regions with same $m$ as the EW baseline but larger $\ell$. As we noted earlier, 
in this case the visibilities of the two diagonal baselines are exactly complex conjugates, so in the $\mathbf{R}(\ell,m)$ matrix
they occupy the same region. 

\begin{figure}
\centering
\includegraphics [width=0.45\textwidth] {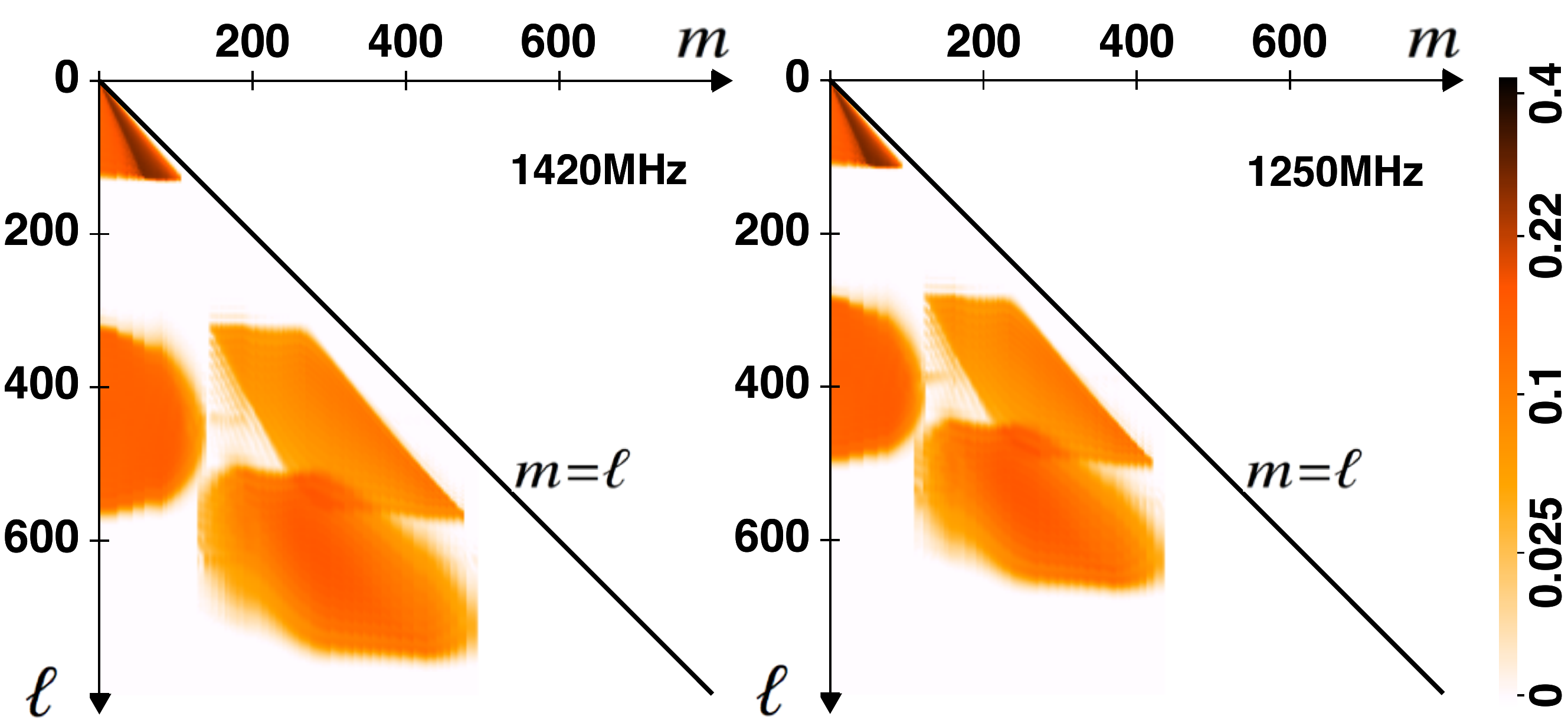}
\caption{The $\mathbf{R}(\ell,m)$ for 4 dishes with 15~m baselines observing at 1420 MHz (left) and 1250 MHz (right) for the 
survey of the region with $35^\circ 23'^\circ<\delta<59^\circ 23'^\circ$. }
\label{fig-4dishlongbase}
\end{figure}

Having discussed the \rzrefreqa{core} response matrix for the two scan case, we turn now our attention to the case of the 
survey of a continuous band from $\delta_1$ to $\delta_2$ with the same array configuration
($2 \times 2$ and $d_{\rm sep}=15 \mathrm{m}$). 
The corresponding $\mathbf{R}(\ell,m)$ matrices for two observing frequencies, 1420 MHz and 1250 MHz,
are shown in Fig.~\ref{fig-4dishlongbase}, with  linear  scales. With the wider band of sky, the $\mathbf{R}(\ell,m)$ matrix 
can be regarded {\rzrefreq approximately} 
as the superposition of the  individual narrow strips, \rzrefreqa{the $(\ell,m)$ plane coverage} by individual baselines are 
distinctly seen. We also note the region due to autocorrelation shows a ``highlighted'' region between $m=\ell \cos  \delta_1$
and $m=\ell \cos  \delta_2$, but within the region $m < \ell \cos  \delta_2$ the value is also non-zero due to the 
superposition, {\rzrefreq and for the same reason the ``inner boundary" at $m=\ell \cos \delta_2$ 
is less clear cut than the ``outer boundary'' at $m=\ell \cos\delta_1$.} At the two frequencies, 
the general shape of the $\mathbf{R}$ matrices are similar, but shifted in position. This is expected,
as the $(\ell,m)$ individual beam positions and extensions vary as $1/\lambda$. However, with a 
separation distance $d_{\rm sep} \sim 3 D_{\rm eff}$, there are large uncovered regions in the $(\ell,m)$ plane,  
and the number of $(\ell,m)$  modes simultaneously measured at different frequencies would be smaller
 than a configuration with fully covered $(\ell,m)$ plane, which would lead to stronger beam frequency dependence and mode mixing
in 21--cm intensity mapping observations.

\subsection{PAON-4 beam and $(\ell,m)$ plane response}

Traditionally, interferometer arrays are employed to achieve high angular resolution, which requires long baselines. 
However, as shown above, with long baselines  there are inevitably {\em holes} on the $\uv$ or $(\ell,m)$ plane 
which will not be covered during observations, as demonstrated in 
Fig.~\ref{fig-4dishlongbase}, where the shortest baselines are 15 m. For sparse images, e.g. a sky dominated by point sources, 
good image reconstruction may still be achievable. However, for reconstructing the diffuse intensity distribution such as the 21--cm 
signal, this will be a major obstacle, as the missing or unobserved modes would be different at different frequencies, 
making it hard to separate the cosmological 21--cm signal from the strong continuum foreground. 
If the baselines are sufficiently short, then at least within certain spatial frequency ranges, the $(\ell,m)$ plane sampling would be
complete, and a better sky reconstruction becomes possible. That's why we shall consider more compact arrays below. 

\begin{figure*}
\centering
\includegraphics [width=0.8\textwidth] {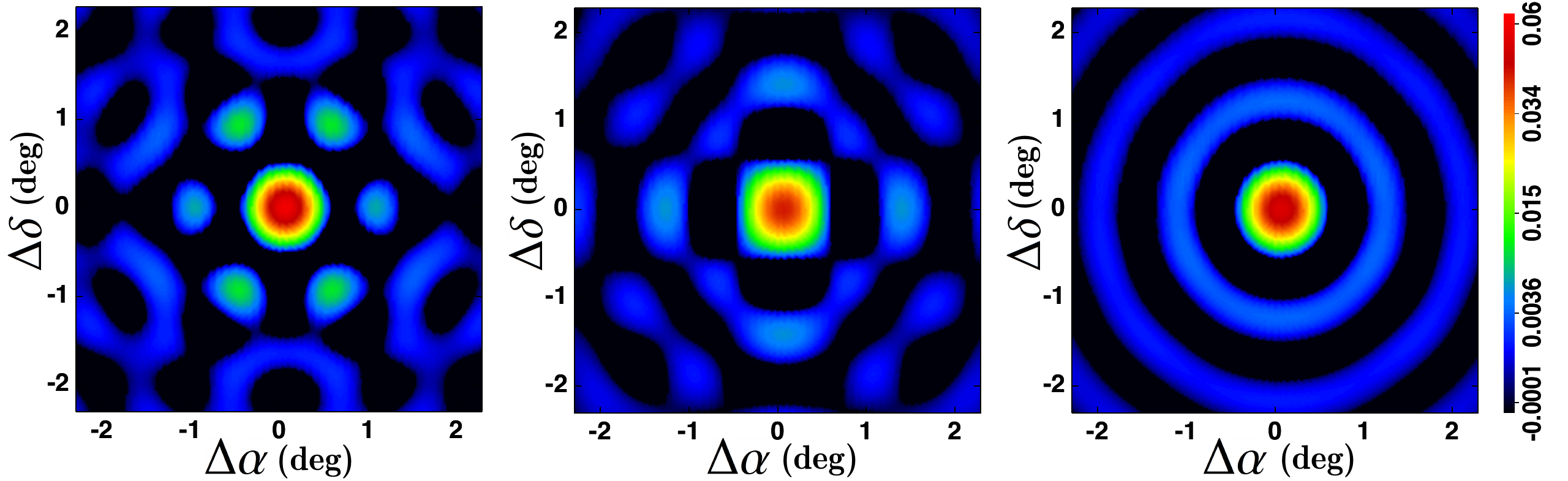}
\caption{Comparison of the PAON-4 beam (left panel) with  that of the compact $2 \times 2$ array (centre panel) and the 
D=15.5 m single dish (right panel).$\Delta \alpha$ and $\Delta \delta$ are the right ascension and declination difference relative to the 
centre. The colour-scale should be interpreted as the ratio of the reconstructed pixel values to the single pixel value in the input map 
representing the point source. }  
\label{fig-paon4-beams}
\end{figure*}

\begin{figure*}
\centering
\includegraphics [width=0.8\textwidth] {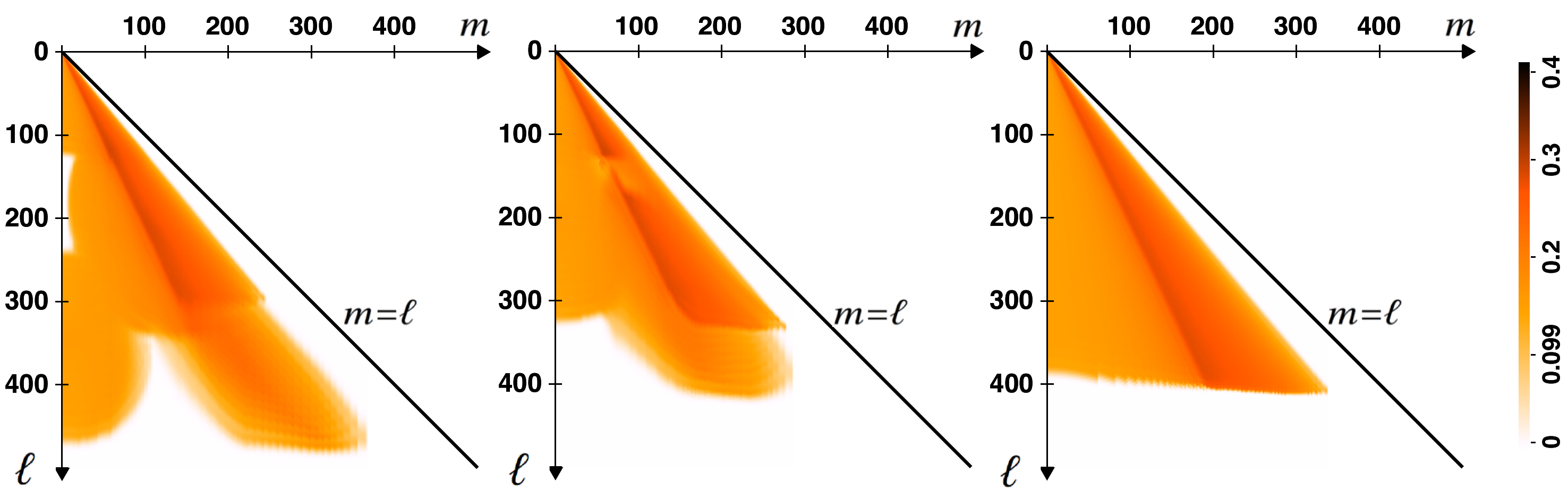}
\caption{Comparison of $\mathbf{R}$ matrix for PAON-4 (left), $2 \times 2$ (centre) and D=15.5 m single dish 
$(\mathrm{D_\mathrm{eff}=14 \, m})$ configuration (right) }
\label{fig-BA-paon4}
\end{figure*}

%
{\rzrefreq
With these considerations we choose the small separations between the PAON-4 dishes, which are 
only slightly longer than the closest-packed configuration. Below
we also compare it with a compact $2 \times2$ regular array with $d_{\rm sep}=7$m, and a 15.5 m diameter 
single dish $(\mathrm{D_{\rm eff}=14 \, m}$) in order to understand its performance.
We consider  a survey composed of 25 constant declination scans, a total of \mbox{$25 \times 6=150$} cross correlations 
are used for the map making, compared to \mbox{$25 \times 4=100$} for the $2 \times 2$ case. 
For comparison, the survey for the large single dish is assumed to be made of 79 constant 
declination scans.} We calculate the beam, or the response to a point source from the full sky reconstruction, as described in 
Sec.~\ref{sec-synbeam} for the PAON-4, $2\times 2$ and large single dish  configurations. The reconstructed beam 
depends slightly on the source declination,  the beams shown here correspond to the central declination of 
observation, i.e. $47^\circ$ N.
Figure~\ref{fig-paon4-beams} shows the beams for the PAON-4 case (left panel) compared with the 
the $2 \times 2$ array (centre panel) and the large single dish (right panel). One can see that the PAON-4 beam has a hexagonal 
symmetry, generated from the product of triangular symmetry and reflection symmetry, which is much more circular in shape than 
the $2 \times 2$, and its resolution is also slightly better than the $D_{\rm eff}=14 \, \mathrm{m}$ single dish.

Figure \ref{fig-BA-paon4} shows the $\mathbf{R}(\ell,m)$ matrix for PAON-4 (left), 
compared with the compact $2 \times 2$ array (centre) and the single dish (right) configurations.  
In Sec.\ref{sec-transitmapmaking} we have already 
discussed the behaviour  of the $\mathbf{R}$ matrix for individual baseline and for individual pointing, as well as the cases of
an $2\times 2$ array with longer baselines. With insight gained from that exercise, we can analyse the  
$\mathbf{R}$ matrices here, which to a good approximation is the linear superposition of different baselines and individual pointing directions. 
The $\mathbf{R}$ matrix for the single dish is very simple, which has a triangular shape. 
It extends to  $\ell_{\max}=2\pi D_{\rm eff}/\lambda \approx 420 $ in the present case 
but otherwise is similar to the auto-correlation for the small dish discussed before.  The triangle is bounded by 
$m=\ell \cos\delta_1~ \approx 0.83 \ell $, and inside the triangle there is an inner boundary at $m=\ell\cos\delta_2  \approx 0.51 \ell$, 
and the $\mathbf{R}(\ell,m)$ is largest between the two boundaries. 
However, the superposition of the stripes also fill up the region with $m<\ell\cos\delta_2  \approx 0.51 \ell$, which is understandable, as 
the modes in this part of $(\ell,m)$ space are basically modes along the NS direction, for which the information is available with 
the superposition of many narrow strips along the latitude, but would not be available for a single narrow strip. 
Compared with the non-compact $2\times 2$ array with longer (15~m) baselines, the compact $2\times 2$ array considered here 
has shorter baselines, so that the region for the NS and EW baselines overlap with each other and also with the auto-correlation part. 
The diagonal baselines are also there, with $m$ similar to the EW baseline but with larger $\ell$.  Thanks to the overlap, now the $(\ell,m)$ 
modes up to certain $(\ell_{\rm max}, m_{\rm max})$ can be measured completely with this array configuration. 
For the PAON-4 array, there are six independent baselines, so the coverage in the $(\ell,m)$ space is more complicated, but generally 
the regions are better covered, and actually extend to higher $\ell$ values and larger area in the $(\ell,m)$ plane.
The little hole in the region $(150<\ell<250, m \sim 0)$ is due to the lack 
of a short north-south baseline with $d \sim 6 \, \mathrm{m}$, and the uncovered area around $(\ell \sim 400, m \sim 120)$ 
is due the lack of a baseline with $d_{\rm NS} \sim d_{\rm EW} \sim 6 \mathrm{m}$.

%

\subsection{PAON-4 noise power spectrum and transfer function }
\label{paon4noise}

\begin{figure*}
\centering
\includegraphics [width=0.8\textwidth] {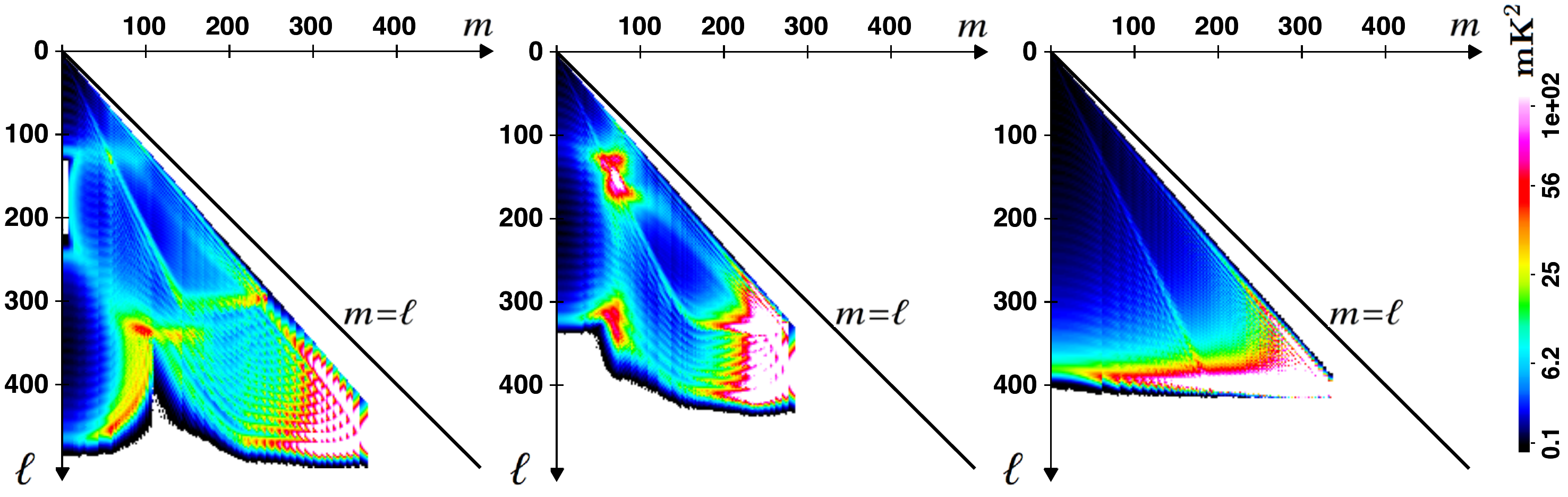}
\caption{Comparison of the error variance matrix for PAON-4 (left), $2 \times 2$ (centre) and D=15.5 m single dish $(\mathrm{D_\mathrm{eff}=14 \, m})$ configuration (right). The scale indicated by the colour-bar corresponds to $\sigma_\mathrm{noise} = 1 \, \mathrm{mK}$.  }
\label{fig-errorvar-paon4}
\end{figure*}

The auto-correlation signals are usually not used in interferometric observations. Indeed, these
signals are very sensitive to the variation of noise level, which can easily swamp the sky signal. 
Only for the case of white stationary noise,  the noise term in the auto-correlation  contribution can be subtracted.  
For cross-correlations, the average contribution remains zero unless there is a correlated noise source between the two different receivers:
$ \langle n_i \, n_i \rangle \propto T_{\rm sys}^2;  \, \langle n_i \, n_j \rangle = 0, \, i \ne j.$
{\rzrefreq But without the auto-correlations, low spatial frequency modes are not sampled, degrading inevitably 
the reconstruction for small $(\ell, m)$ values.} Using small separation between
the dishes will help reduce the unobserved modes in the low $(\ell,m)$ region. This is also one of the 
reasons why we use small dishes and close packed array configurations, which also avoids the 
incomplete coverage at higher $(\ell,m)$. 
{\rzrefreq Below we shall assume that the reconstruction is performed only with 
cross-correlations for the interferometer arrays.}


%


The noise level for visibility time samples are computed according to Eqs. (\ref{eq-signoise-1})-(\ref{eq-signoise-2}). 
For the reconstructed maps discussed here we have chosen $\ell_{\rm max}=1500$ and HEALPix $\mathrm{n_{side}}=512$, 
thus we have $n_t =  2 m_{\rm max}=3000$ visibility time samples over 24 hours. 
For a total survey duration of $t_{\rm survey} = 175$ days ($\sim$ 6 months), each $\mathrm{\delta-scan}$ would 
be repeated 7 times, leading to a total integration time of $t_{\rm int} \sim 201.6 \, \mathrm{s}$ per time sample. 
Assuming a system temperature $T_{\rm sys} = 100 \, \mathrm{K}$ and $\Delta \nu=1 \, \mathrm{MHz}$ frequency bin
width, the noise level is $9.96 \, \mathrm{mK}$ \rzrefreqa{for each visibility time sample,} for the PAON-4 and $2 \times 2$ cases.  
For the single dish case, each of the 79 scans would be repeated twice, leading to a total survey 
duration of 158 days, slightly shorter than the PAON-4 case. 
Each time sample would then have a total integration time of $\sim 58 \, \mathrm{s}$, leading to a 
noise level  $\sim 18 \, \mathrm{mK}$. 

Figure \ref{fig-errorvar-paon4} shows the error covariance matrix, defined in Eq.~(\ref{eq-covardiag}),
for the PAON-4 (left), $2\times 2$ (middle)
and 14 m single dish (right) survey. In these maps, we have set a very large error for all points with no data at all, 
so that the regions which are well-measured (low noise)  are represented by dark colour. We can see that
 there are some similarities in the distribution with the $\mathbf{R}(\ell,m)$ matrix, but 
unlike the $\mathbf{R}$ matrix distribution, which is fairly smooth, here we see a lot more variations in the distribution, as weights
also goes into the number of redundant baselines.  The single dish survey is still the simplest, it achieves uniformly low-noise measurement 
for $\ell<300$, but the error blows up at $\ell>350$. The 
$(\ell, m)$ modes at $m >  \ell \cos \delta_2$ are measured primarily along the EW direction within the survey region, and 
those modes with $m <  \ell \cos \delta_2$ are measured primarily along the NS direction. Interestingly, there is a fairly 
large error at the border line $m = \ell \cos \delta_2$, \rzrefreqa{which might be related to evolution of the diagonal 
band structure with $\ell$, in each $\mathbf{R}_m$ matrix. }

\begin{figure}
\centering
\includegraphics [width=0.4\textwidth] {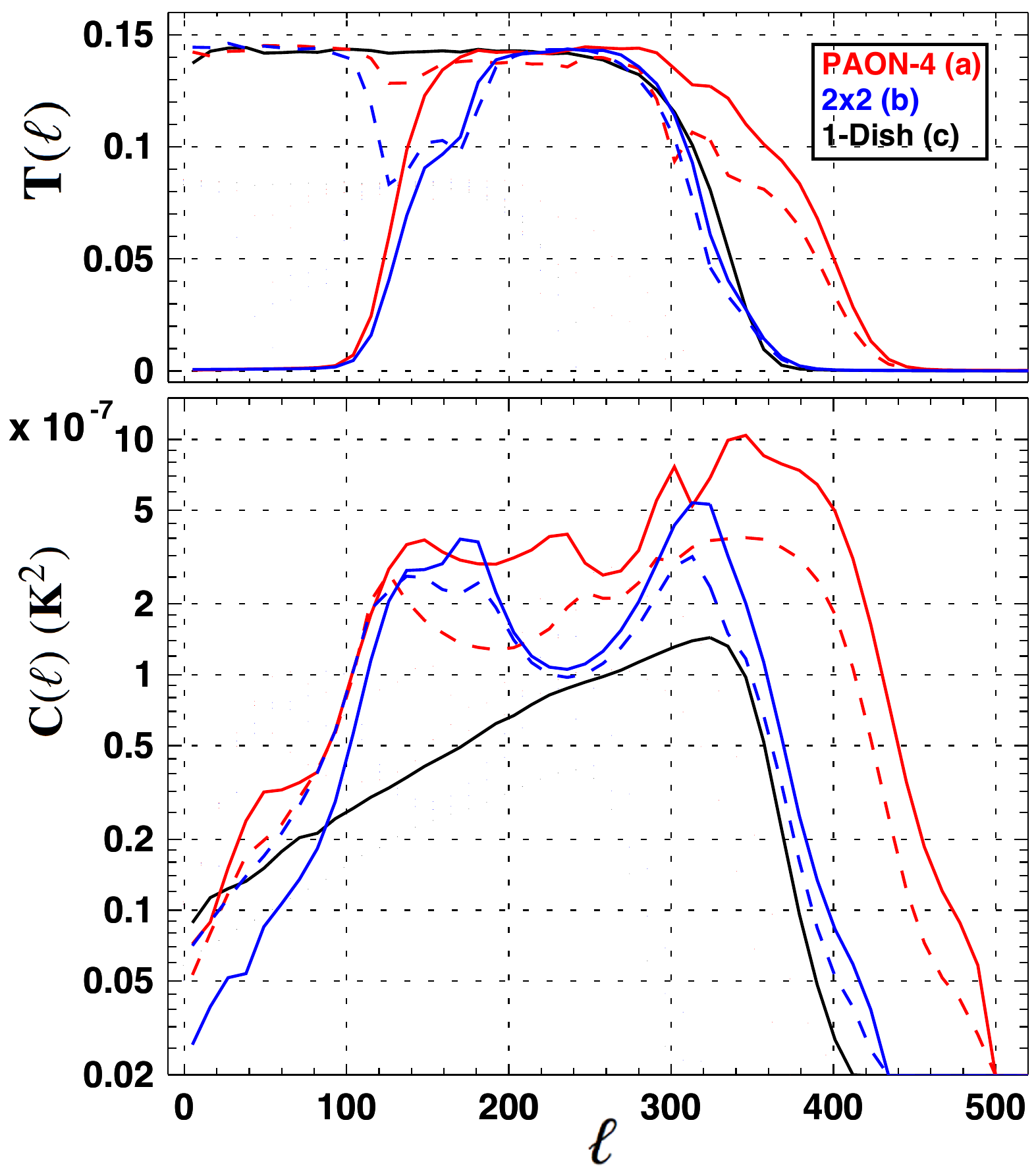}
\caption{Comparison of the transfer function $T(\ell)$ (top) and noise power spectrum 
$C^\mathrm{noise}(\ell)$ (bottom) for PAON-4 in red, compact $2 \times 2$ array in blue, and the single dish in black.
The dashed and solid lines show respectively the results with and without auto-correlation.  }
\label{fig-clnoise-tl-gb-paon4}
\end{figure}

For the two arrays, the distribution is fairly complicated, though we can see where the $\mathbf{R}$ matrix is substantial, the
measurement error is also relatively small.  The region around the centre $(\ell,m)$ value for each baseline is a basin of low measurement error, 
separated by some watersheds with larger errors. This is easy to  see in the case of the 4 baselines of the $2\times 2$ array (recall that the 
two diagonal baselines coincides each other on the $(\ell, m)$ map, when restricted to $m\geq 0$ half plane). 
For the case of the PAON-4 array there are six independent baselines, so the distribution is even more complicated but can also be identified. As 
in the single dish case, along the watershed at 
$m = \ell \cos \delta_2$ the error is somewhat larger. The interferometers do probe higher $\ell$ region, up to $\ell \sim 500$ in the case of 
PAON-4, but there are also some regions with no measurement. The small value at the edge of large $\ell$ (bottom of the figure) is however an 
artefact of the computation procedure: as the elements of the $\mathbf{L}$ matrix goes to zero, the corresponding elements in the pseudo-
inverse  $\mathbf{H}$ are also set to zero, and as a result we see the response function and error drop to small values at the edges, but this 
does not affect our final estimation of the measurement error.

As discussed in section \ref{sec-lm-filtering}, additional filtering and masking can be applied after the 
determination of the sky spherical harmonic coefficient to improve the result. In addition to the inverse variance weighted 
filter $W_1$, we can also mask out
the border pixels outside of the band $34^\circ 23' < \delta < 60^\circ 23'$, and also apply  
a smoothing filter function $W_2(\ell)$ with $\ell_{\rm max} \sim 420$ to suppress modes with high noises. 
{\rzrefreq The map pixel resolution for HEALPix $\mathrm{n_{side}}=256$ used here is $\sim 0.25^\circ$. The reconstructed map pixel values
dispersion (RMS) due to noise on visibilities decreases from $146 \, \mathrm{mK}$ to  $55 \, \mathrm{mK}$ after the filtering,
at the expense of reducing the  $(\ell,m)$ coverage somewhat. 
}

In Fig.\ref{fig-clnoise-tl-gb-paon4},  we plot the angular power transfer functions $T(\ell)$ (top panel) and 
the noise angular power spectrum $C^\mathrm{noise}(\ell)$ (bottom panel) for the PAON-4 array, the $2\times 2$ array and the large single dish. 
As expected from the map-making algorithm used here, for the single dish the transfer function is nearly 
constant with respect to $\ell$, up to some $\ell_{\rm max}$ between 300 and 400, above which it drops precipitously.  
The value of the plateau region in the transfer function ($\sim 0.16$)
is determined by the fraction of mapped sky $f_{\rm sky}$, which is given by
$ f_{\rm sky} = \frac{1}{2} \left( \cos \delta_1 - \cos \delta_2 \right) $.  
{\rzrefreq
The noise power spectrum increases smoothly
with angular frequency ($\ell$), up to $\ell=400$, at which point it drops as the transfer function vanishes.
For the arrays, without the auto-correlation (solid lines), 
the transfer function is nearly zero at small $\ell$, but increases to up to the plateau value 
at around  $\ell = 150$. If we do include the auto-correlations (dashed lines), then the plateau would extend all the way to $\ell=0$.  
For the $2 \times 2$ configuration, there is a dip around $\ell = 150$, due to the 
high noise modes around the $(\ell=150, m=100)$ visible in the error variance matrix, for PAON-4 the dip is smaller and the 
whole transfer function is smoother. The noise power spectrum drops to very small value at 
 $\ell \lesssim 100  ~(\theta \gtrsim 2^\circ)$. } 
Again, the PAON-4 and $2 \times 2$ configurations present 
some structures in  the noise power spectrum $C^\mathrm{noise}(\ell)$ as in the transfer function. 
However, the $C^\mathrm{noise}(\ell)$ curve for PAON-4 is smoother than that of the $2 \times 2$ (b) configuration.

\section{Application to the Tianlai Dish Array}
\label{sec-tianlaicase}

The method and analysis criteria presented above can also be applied to the Tianlai experiment, 
and the insight gained from the analysis of the PAON-4 array in the previous section would be useful for the 
understanding the bigger Tianlai 16-dish array response. 
As in the PAON-4 case, we also have studied a number of configurations but we will focus here only on two configurations, 
the square $4 \times 4$ layout and a circular array layout, the latter (circular) being the current Tianlai configuration
16-dish array configuration. As discussed in Sec. \ref{sec-paon4respmtx}, we shall only consider compact dish array layouts to  
ensure a complete sampling of the $\uv$ plane or  $(\ell,m)$ space. 

%

\subsection{Blockage factor, beams, and response}
\begin{figure}
\centering
\includegraphics [width=0.35\textwidth] {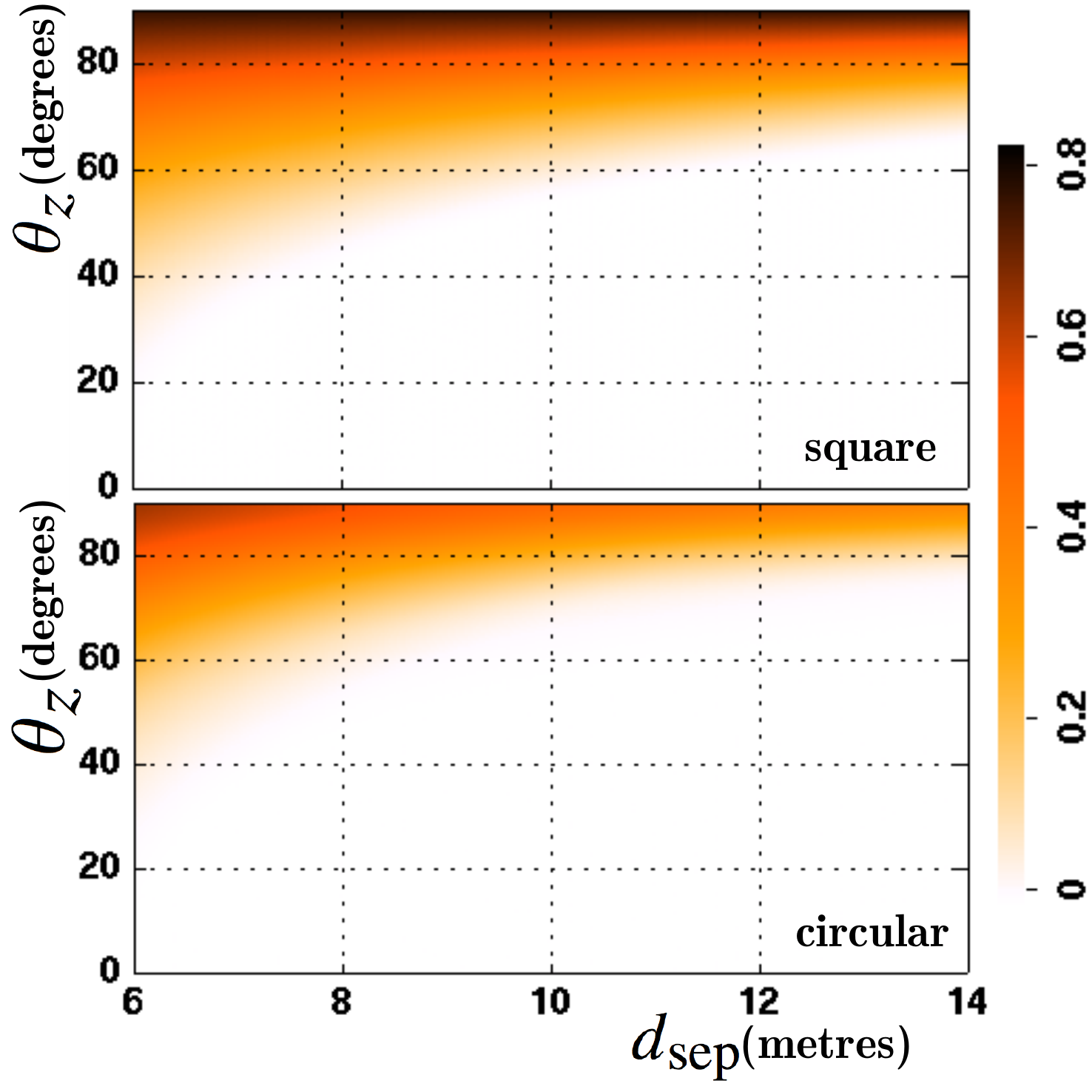}
\caption{The blocking factor for the square $4 \times 4$ configuration (top) and the circular configuration (bottom)  
as a function of separation scale $d_\mathrm{sep}$ and antenna pointing angle with respect to the zenith $\theta_Z$. The colour bar
shows the blocking factor.}
\label{fig-blocking}
\end{figure}

Before discussing the beam pattern and map-making capability, we first consider the blockage of the antenna with different 
separation scale $d_\mathrm{sep}$. For simplification, we treat the antennas as circular dishes of 6-m diameter
placed on the same horizontal plane, aligned in the North-South direction, and  then we define the geometric blocking factor, 
which is the overlapped projected cross section divided by the total area of the dishes. This ignores the effect of diffraction and multiple
overlaps, but is easy to compute. The blocking factor as a function of separation scale $d_\mathrm{\rm sep}$ and zenith angle $\theta_Z$
are shown in Fig.\ref{fig-blocking} for the square (top) and circular (bottom) arrays. 
\rzrefreqa{  The square array has generally higher blocking factor, 
as, contrary to the circular array, the antennas are aligned along the north-south direction. }
Obviously at small zenith angle, the dishes will not block each other whatever 
the  $d_\mathrm{sep}$ value is. As the zenith angle increases, dishes may be partially blocked by their neighbours 
to the north or south, though for each dish the blocking is different. 
{\rzrefreq For the circular configuration with the minimum 
6 m separation, blocking factor $< 10\%$ is only achievable at zenith angles $\theta_Z< 47^\circ$. 
However, the pointing range is extended with increasing $d_\mathrm{sep}$ increases for the same blocking factor. 
For example, at  $d_\mathrm{sep}=9$m, we read from the figure that blocking factor is less than $<10\%$  for 
zenith angles up to $70^\circ$, }
which is more or less the maximum zenith angle observation foreseen with Tianlai.  
To minimize the ground preparation work, we have tried a few different values of $d_\mathrm{sep}$ around 9-m and 
calculated the positions of the antennas in the circular array configuration. Finally the value $d_\mathrm{sep}=8.8 \, \mathrm{m}$ 
was chosen which correspond to the current configuration.  

\begin{figure*}
\centering
\includegraphics [width=0.8\textwidth] {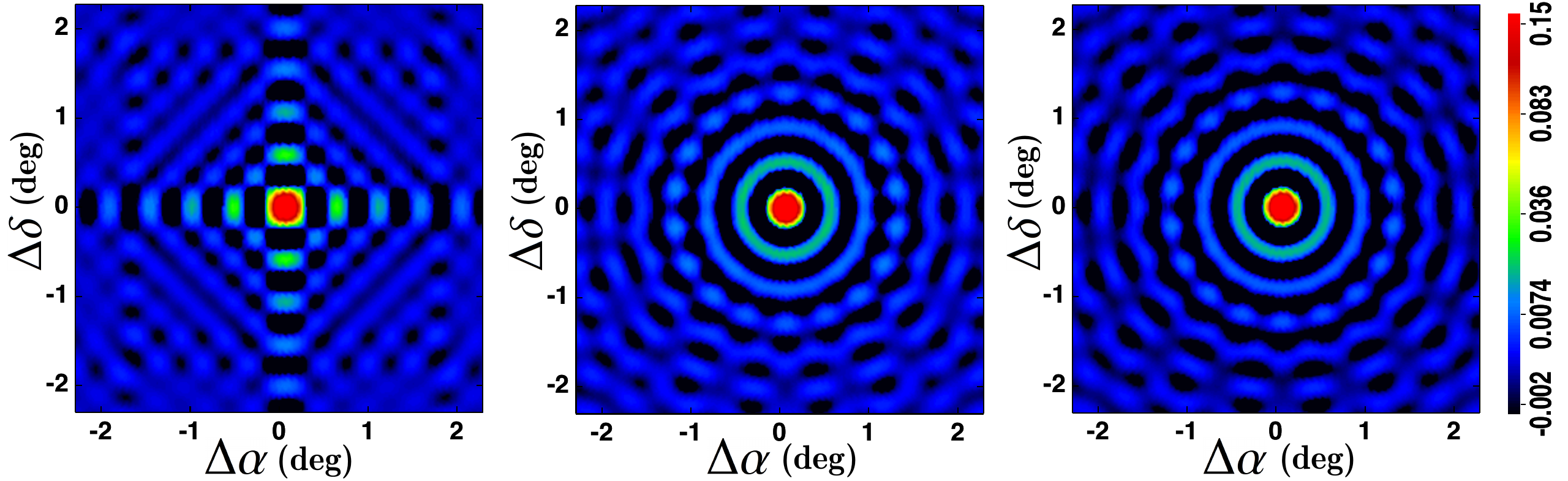}
\caption{Comparison of Tianlai 16-dish circular array with auto-correlation (centre panel) with a square $4 \times 4$ array (left panel). 
The beam for the Tianlai 16-dish circular array, without auto-correlation signals is shown on the right panel. 
$4.4 \times 4.4 \, \mathrm{deg^2}$ high resolution area extracted from the reconstructed maps, centred on a point source position. 
The colour-scale should be interpreted as the ratio of the reconstructed pixel values to the single pixel value in the input map 
representing the point source. }
\label{fig-d16-beams}
\end{figure*}

Fig. \ref{fig-d16-beams} shows the reconstructed 2D synthesized beams, i.e. the reconstructed map for a point source, 
for the square array (left panel), the circular array with 
auto-correlation included (Centre panel) and the circular array without including the auto-correlation (right panel). 
The square array exhibits a strong cross-shaped grating type pattern in its beam, while for the circular array,
the beam is nearly circular-symmetric. The beam formed with and without auto-correlation signals are similar, 
with only subtle differences on large scales, so that using only the cross-correlation signals would not much affect the 
observations except on very large angular scales.

\begin{figure*}
\centering
\includegraphics [width=0.9\textwidth] {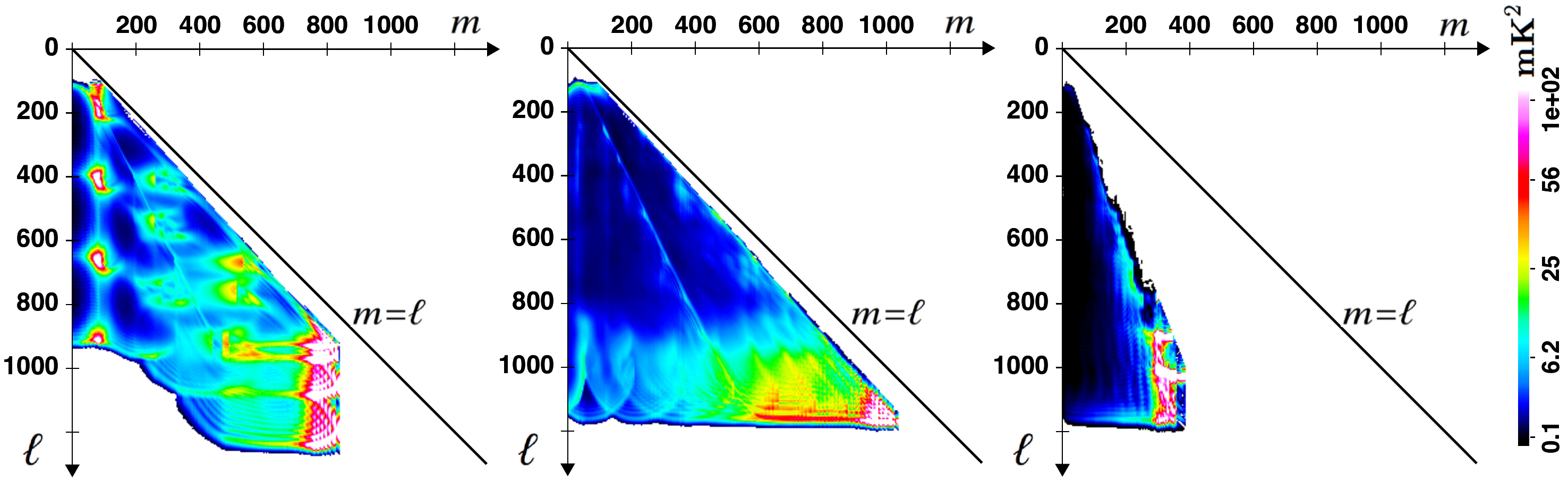}
\caption{The error variance matrix for 16 dishes array surveys. Left: the square $4\times 4$ array mid-latitude survey; 
Centre:  the Tianlai 16-dish circular array mid-latitude survey; Right: the Tianlai 16-dish circular array polar cap survey.
The scale indicated by the colour-bar corresponds to $\sigma_\mathrm{noise} = 1 \, \mathrm{mK}$ }
\label{fig-errorvar-d16}
\end{figure*}


\begin{figure}
\centering
\includegraphics [width=0.35\textwidth] {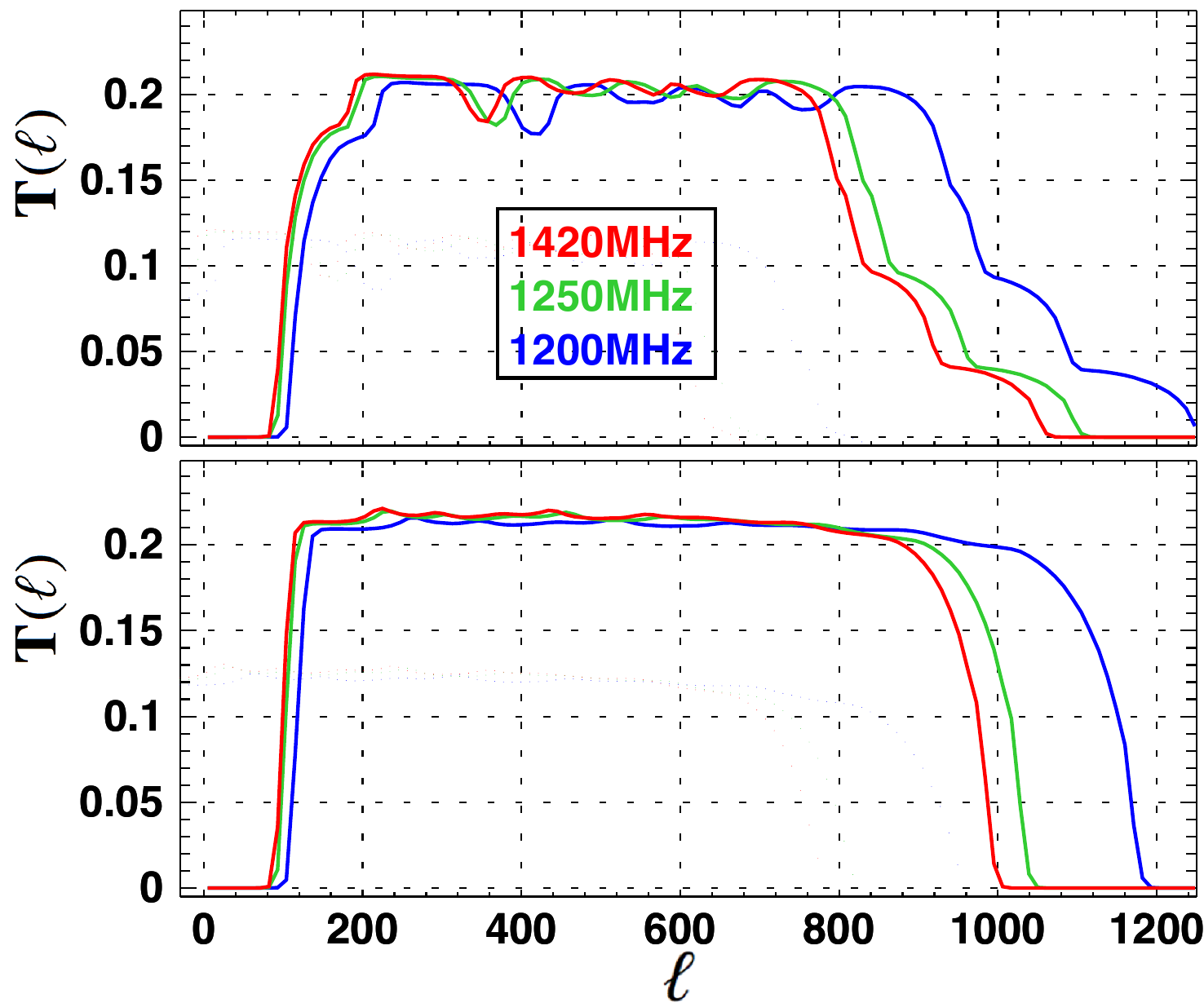}
\caption{The transfer function for the 16-dish square (top) and circular (bottom) configurations. The curves corresponds 
to three observations frequencies: 1420 MHz (blue), 1250 MHz (green), 1200 MHz(red). } 
\label{fig-tf16}
\end{figure}

{\rzrefreq
We shall consider two surveys for the Tianlai array. One is a mid-latitude survey, at the latitude of the Tianlai array site
this is achieved by constant elevation scans around the zenith. We shall assume that the survey consists of 
31 constant $\delta$ scans, each repeated 7 times, corresponding to a total survey duration of $31\times7=217$ days.
The total integration time for each visibility time sample would be $t_{\rm int} \sim 201.6 \, \mathrm{s}$ 
for $n_t=2 m_{\rm max}=3000$. Using Eq.(\ref{eq-signoise-1}) and Eq.(\ref{eq-signoise-2}), and  
assuming a system temperature $T_{\rm sys} = 100 \, \mathrm{K}$, and $\Delta \nu=1 \, \mathrm{MHz}$, 
we obtain a visibility noise level of $\sigma_{\rm noise} \sim 9.96 \, \mathrm{mK}$ per visibility time sample, similar to the PAON-4 case.
}

\begin{figure}
\centering
\includegraphics [width=0.35\textwidth] {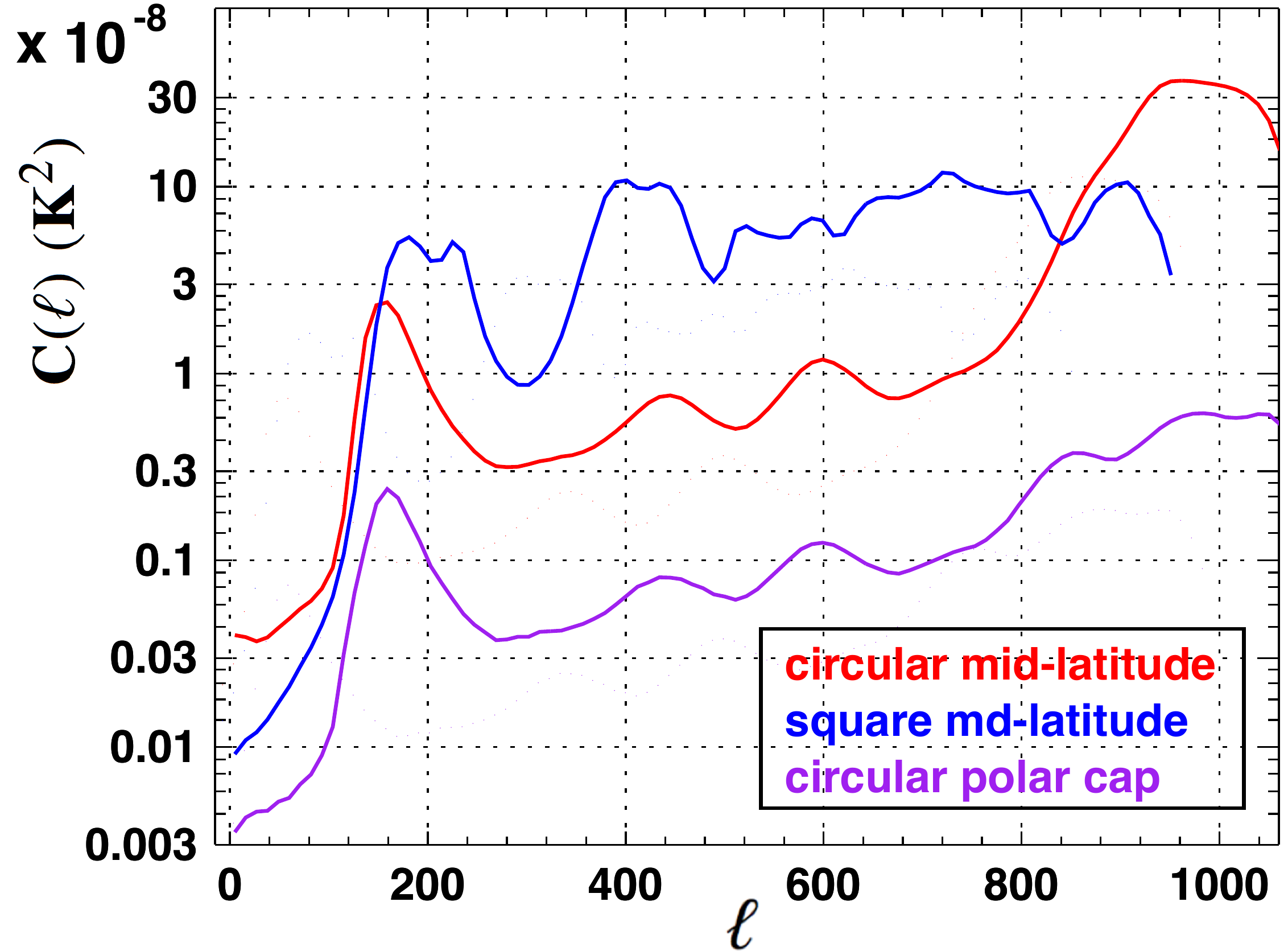}
\caption{Comparison of the noise power spectrum $C^\mathrm{noise}(\ell)$ at 1420 MHz. The blue and red curves correspond respectively 
to the square and circular array mid-latitude surveys, and the purple curve to  the circular array polar cap survey. } 
\label{fig-noisecl-d16}
\end{figure}

{\rzrefreq The second observation programme with the Tianlai dish array considered here is the polar cap survey. 
The high latitude or polar cap region is both scientifically interesting and have some specificities in terms of observation and processing. 
From the perspective of an observer on the ground, the sky rotates around the celestial pole, if one points 
the telescope to the pole, the same point will be observed all times, so that much deeper exposure of 
this small region can be achieved within a relatively short time. It might therefore prove interesting to carry out 
the observation first in the polar cap region. The polar cap consists of 16 scans, each shifted by $1^\circ$ in declination,
starting from the north celestial pole ($\delta = 90^\circ$), down to $\delta=75^\circ$, leading to observation of the sky region
$75^\circ \lesssim \delta < 90^\circ$. 
We assume a total survey duration comparable with the mid-latitude case, though the covered declination range is 
about half of the mid-latitude survey. Each declination will then be observed 14 times, twice the number of 
mid-latitude case, requiring $16\times14=224$ days, about 7.5 months to carry the full survey. 
The actual sky area is even smaller at such high latitudes ($\sim 600 \, \mathrm{deg^2}$), compared with
$\sim 7000 \, \mathrm{deg^2}$ for the mid-latitude survey, so that we expect a much deeper survey, with reconstructed map noise 
level around $\sim 3$ times lower for the polar cap survey.  We use here higher resolution HEALPix maps 
with $\mathrm{n_{\rm side}} = 1024$ to minimise distortions near the pole.  
The integration time per sample would then be  $t_{\rm int} \sim 403.2s$, and assuming a system temperature 
of $T_{\rm sys} = 100 \, \mathrm{K}$, and $\Delta \nu=1 \, \mathrm{MHz}$, we obtain a visibility noise level of 
$\sigma_{\rm noise} \sim 7 \, \mathrm{mK}$ 
using Eq.(\ref{eq-signoise-1}) and Eq.(\ref{eq-signoise-2}), $\sqrt{2}$ times lower than the mid latitude case. }

\begin{figure*}
\centering
\includegraphics [width=0.9\textwidth] {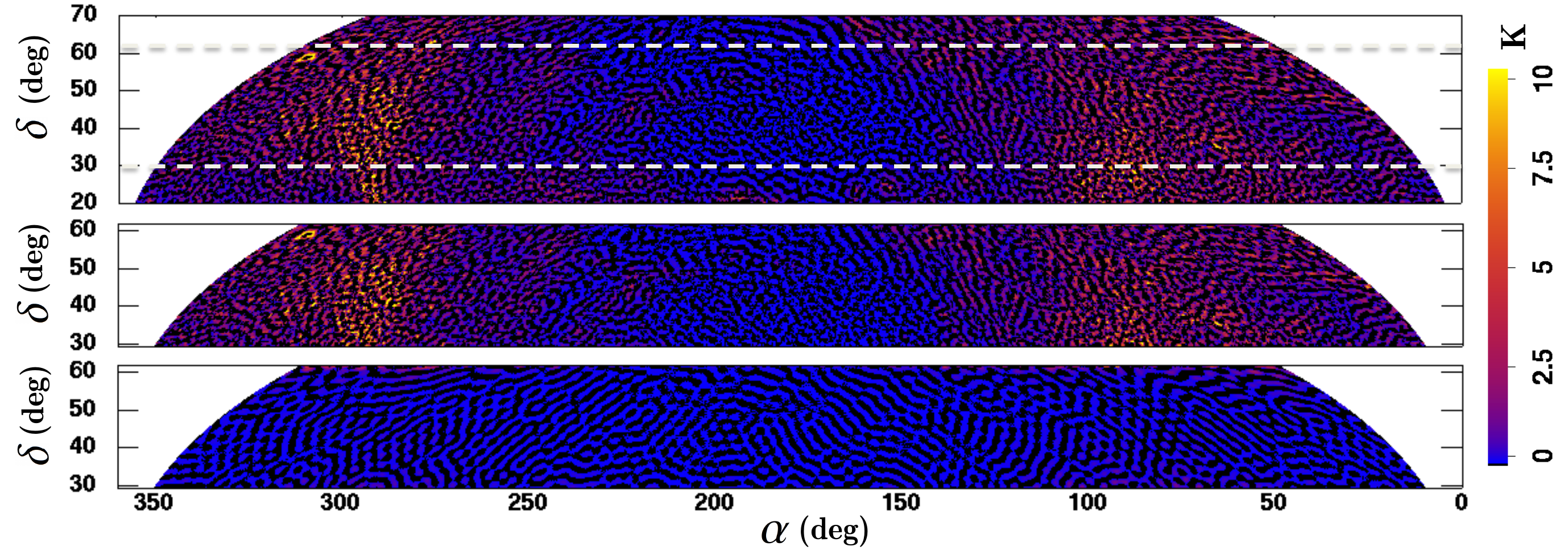}
\caption{Comparison of the reconstructed and the input map for the circular configuration and mid-latitude survey; 
top: LAB map, after application of the high pass $W_3$ filter, in the declination range $20^\circ < \delta < 70^\circ$ (1); 
middle: the reconstructed map in the range $30^\circ < \delta < 60^\circ$ with both $W_1$ and $W_2$ filters (2); 
bottom: the difference map (1)-(2) between (1) and (2) in the range $30^\circ < \delta < 60^\circ$. }
\label{fig-recmapna-d16}
\end{figure*}

\begin{figure*}
\centering
\includegraphics[width=0.9\textwidth]{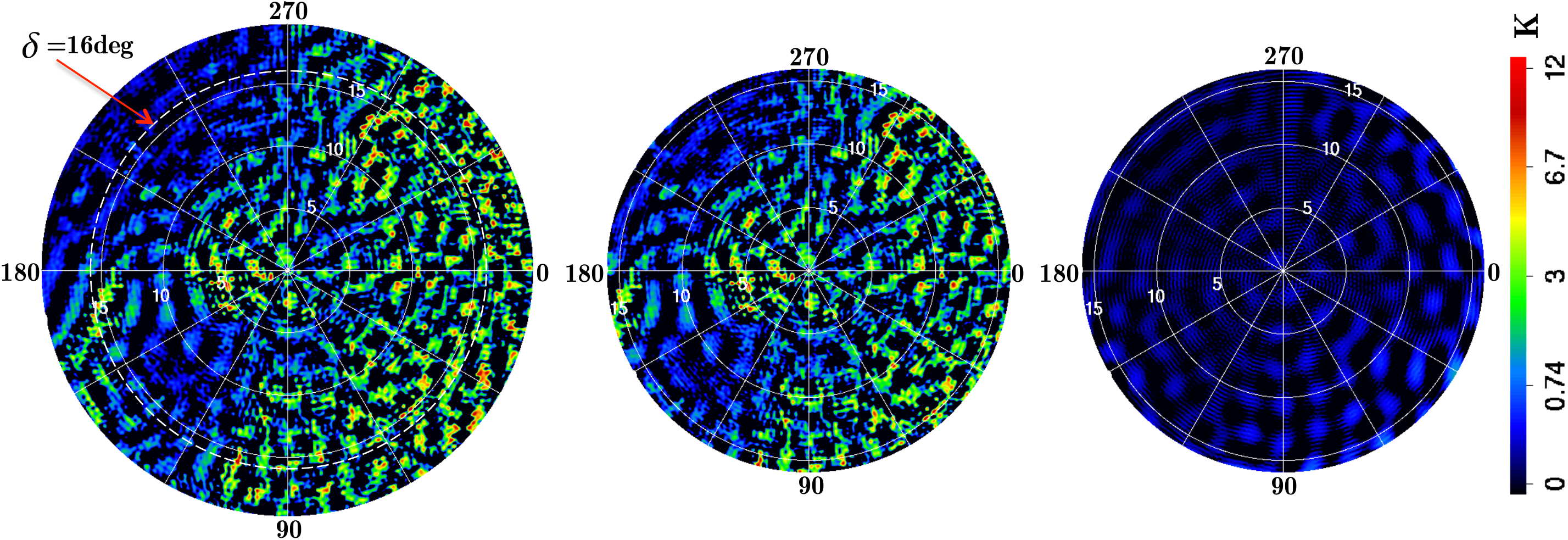}
\caption{ Comparison of the reconstructed and the input map for the circular configuration-polar cap survey;
left: input LAB map with $W_3$ high pass filter, in the polar cap area, within $20^\circ$ radius (1);  
centre: the reconstructed map with Tianlai 16-dish circular array polar cap survey within a radius of $16^\circ$ (2); 
right: the difference map (1)-(2) }
\label{fig-polarmap}
\end{figure*}

{\rzrefreq
Figure \ref{fig-errorvar-d16} shows the error variance matrix 
in the $(\ell, m)$ basis for the square array mid-latitude (left), circular array mid-latitude (centre) and circular array polar-cap surveys 
at 1420 MHz. For the two mid-latitude surveys, the triangles are bounded by $m \gtrsim \cos 29.10^\circ \ell \approx 0.87\ell$.
In spite of the increase in the number of baselines, the coverage in the $\uv$ plane or $(\ell,m)$ space is still
not very dense for the square array, so the most prominent features in that case are the grid patterns, 
e.g. at $(\ell,m)=(185, 94), (414, 94), (668, 94), (414, 282), (556, 282), (764, 282)$, where {\em islands} with 
relatively large errors in the $(\ell,m)$ space are located. However, even in this case, we can see that the errors 
are much smaller than the 4-dish case, as there are many more baselines. 
The circular configuration, on the other hand, has a much larger and more uniform {\em sea} of low error  region, 
though at large $\ell$ corresponding to longer baselines, the errors become somewhat larger. 
Notice that we can still see an {\em island chain} along the $ m =\cos 59.10^\circ \ell \approx 0.51 \ell$
line in both cases, but not as strongly peaked as in the PAON-4 case. For the polar cap survey, the 
$(\ell,m)$ space probed is restricted to $m< \ell \cos 75^\circ \approx 0.26 \ell$, 
however, modes up to $\ell \sim 1200$ are still measured. 
This smaller range of $m$ does not hamper the reconstruction of the map near the polar region, 
because here the temperature variations are indeed described by the smaller $m$ modes.  }

In Fig.\ref{fig-tf16}, we plot the transfer function $T(\ell)$ (measured from cross-correlations only) for the mid-latitude survey of the two arrays,
with the blue, green and red curves for 1420 MHz, 1250 MHz and 1200 MHz observations respectively.  
The transfer function is generally flat over the effective $\ell$ range for the arrays, with some slight wiggles. For the circular array the wiggles are even less pronounced than the square array. It should also be noted that the wiggling structure shifts with frequency. These wiggling features can affect the 
BAO power spectrum measurement, and have to be taken into account in the final analysis. If we can determine the transfer function 
exactly, and the error variance is smaller than the signal, it is possible to correct for it, but it
is however highly desirable to select configurations minimising wiggles and non uniformities in 
the response matrix and transfer function when designing the instrument and the survey strategy.


{\rzrefreq
In Fig. \ref{fig-noisecl-d16} we plot the noise angular power spectrum measured from the cross-correlation data at 1420 MHz, for the 
square array mid-latitude survey (blue),  the circular array mid-latitude survey(red), and the circular array polar cap survey (purple), after 
the application of $W_1$ and $W_2$ filters. For the mid-latitude surveys, the 
noise angular power spectrum of the circular configuration is lower than that of the square array, and also smoother. For the circular array,
the polar survey yields lower noise power spectrum than the mid-latitude survey, which is expected as it covers a smaller sky area. 
}

We can see clearly that the circular array has a better $(\ell, m)$ space coverage, resulting in a higher
reconstruction quality, a more uniform transfer function a lower noise power spectrum. 
That is why we have chosen this configuration for the current Tianlai 16-dish array layout. 
\rzrefreqa{ However, regular arrays present some advantages, in particular for the calibration using redundant 
baselines, as discussed for example by \citet{2010MNRAS.408.1029L}.  Other aspects of the data analysis process, 
including array calibration will be addressed in future studies.  }

{\rzrefreq
Figure \ref{fig-recmapna-d16} sows a comparison of the original and reconstructed maps at 1420.4 MHz derived from 
simulated Tianlai 16-dish circular array observation at mid-latitude as defined above, and illustrates the reconstruction quality. 
The reconstructed map looks somewhat different from the original map, as the auto-correlation visibilities have been ignored 
and the large angular scale features are thus not visible. However, we can apply a high pass filter defined in 
section \ref{sec-lm-filtering} to simulate this effect,
$W_3(\ell) = (1 + e^{(\ell_B - \ell)/\Delta \ell_B} )^{-1}$ , where $\ell_B=120 , \Delta \ell = 10$.  In the top panel, we show 
the LAB map filtered with $W_3$, with the dashed lines marking the limits of the observed region.  
The middle panel shows the reconstructed map with both $W_1$ and $W_2$ filters applied, and restricted to the observed area, 
while the difference map is shown in the bottom panel. 
} 

In Fig. \ref{fig-polarmap} we compare the reconstructed map to the high-pass filtered input map for the polar cap survey. 
Again the reconstructed map resembles the high-pass filtered map except at the borders, beyond the surveyed 
region at $\delta \lesssim 75^\circ$. \rzrefreqa{We see that the map reconstruction, transfer function and noise power 
spectrum computation is correctly handled in the polar region, as well as in the mid-latitude, while higher sensitivities 
could be obtained in the polar cap area thanks to longer per pixel integration time.}

\begin{figure}
\centering
\includegraphics [width=0.45\textwidth] {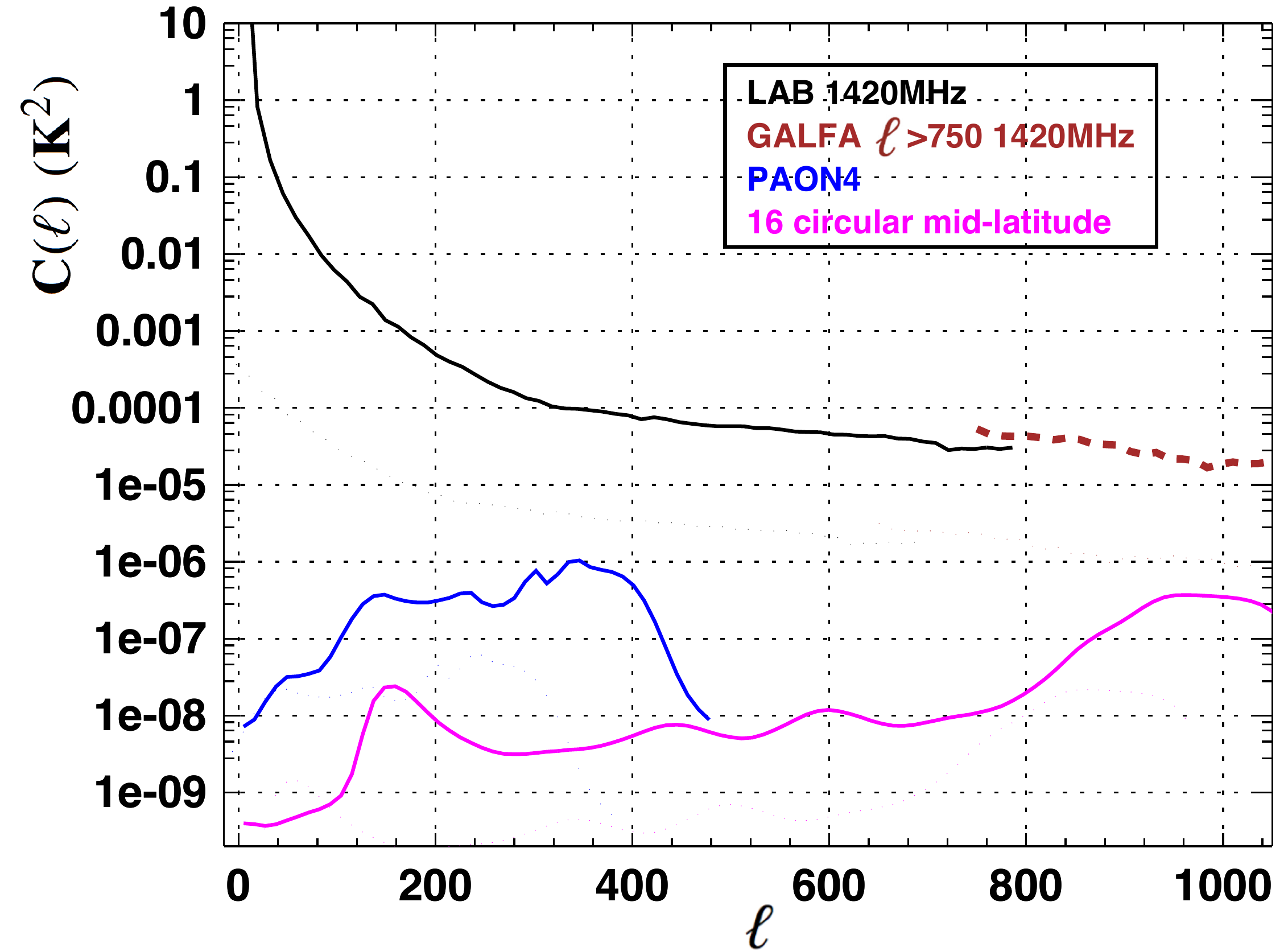}
\caption{Noise power spectrum for the Tianlai mid-latitude survey and PAON-4, compared to Galactic 21--cm signal power spectrum. The power spectrum of the LAB map is shown in black, and the brown dashed curve corresponds to its extension to higher resolution ( $\ell >750$) using GALFA. The blue curve is the noise power spectrum for the PAON-4 case, and the magenta curve corresponds to the Tianlai 16-dish array mid-latitude survey. 
The noise power spectra have been computed without the auto-correlation signals. }
\label{fig-ps-inputmaps}
\end{figure}

\subsection{Tianlai 16-dish array sensitivity}
{\rzrefreq
Figure \ref{fig-ps-inputmaps} shows the comparison of the Milky May 21--cm power spectrum with the 
Tianlai 16-dish and PAON-4 noise power spectra. The 21--cm power spectrum is derived from the 
LAB (for $\ell <750$) survey  and GALFA survey data \citep{2011ApJS..194...20P} survey (for $\ell >750$).
We have rescaled the angular power spectra with sky coverage fraction, 
i.e. what is plotted is $C_\mathrm{map}(\ell) * (1/f_{sky})$, where $C_\mathrm{map}(\ell)$ is the map raw, 
uncorrected angular power spectrum. 
The blue curve in Fig. \ref{fig-ps-inputmaps} is the noise power spectrum for PAON-4,  
and the magenta curve is the noise power spectrum of the Tianlai 16-dish circular array. 
We see that the noise power spectrum for both PAON-4 
and Tianlai 16-dish array are well below the Galactic HI power, so both should be able to measure the 
Galactic HI without difficulty. }

\begin{figure}
\centering
\includegraphics[width=0.45\textwidth]{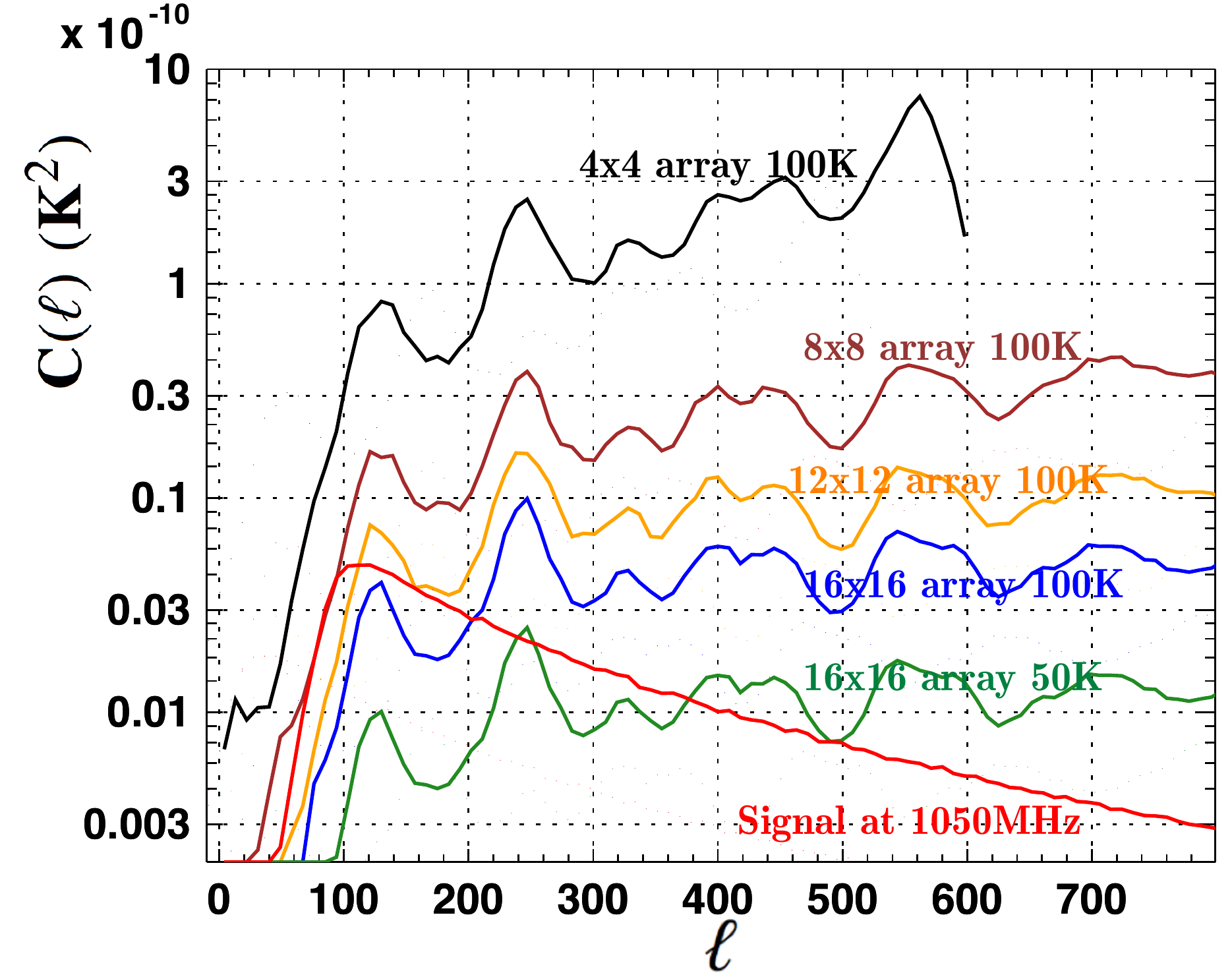}
\caption{The expected angular noise power spectra $C^\mathrm{noise}(\ell)$ at 1050 MHz for a 2 year survey of a $20^\circ$ 
band in declination for square arrays with 16 (black), 64 (brown), 144 (yellow), 256 (blue) dishes with $\Tsys=100 \mathrm{K}$. 
The green curve correspond to the 256 dish array with $\Tsys=50 \mathrm{K}$ and the red curve shows the expected 
cosmological signal at 1050 MHz, filtered by the instrument response $C^{21}(\ell) \times T(\ell)$. }  
\label{fig-arrays}
\end{figure}

Both the PAON-4 and Tianlai-16 dish array are small prototype arrays, 
their sensitivities are not sufficient to detect the neutral hydrogen in the large scale 
structure within reasonable time. 
{\rzrefreq 
In order to detect the latter, arrays with many more elements are required. To see this, 
we compare the noise angular power spectrum with the expected signal from the 
large scale structure $C^\mathrm{sig}(\ell) = C^{21}(\ell) \times T(\ell)$  at 1050 MHz ($z=0.35$), 
taking into account the transfer function $T(\ell)$. We have assumed a global neutral hydrogen relative density 
of $\Omega_{\rm HI} b=0.62 \times 10^{-3} $  \citep{2013MNRAS.434L..46S}  with bias factor $b=1$ to compute the 
expected cosmological 21--cm signal $C^{21}(\ell)$. To simplify the computation and also make the comparison easier,  
we consider several regular square arrays, with 16, 64, 144 and 256 dishes. 
In Fig.\ref{fig-arrays} we plot the forecast noise angular power spectra for these four configurations, 
respectively $4\times4$, $8\times 8$, $12 \times 12$, and $16\times 16$ D=6 m dish arrays. 
We have considered a survey covering a $20^\circ$ band in declination $(35^\circ 10' < \delta < 53^\circ 10')$, 
composed of  13 constant declination scans, each shifted by 1.5 degrees, with a longer total survey duration (2 years).
We have also shown the $C^\mathrm{noise}(\ell)$ for $16 \times 16$ array with a $\Tsys=50 \, \mathrm{K}$, which could 
be considered as the design target. 
We see that the noise angular power spectrum drops steadily with the array size, scaling roughly as $\mathrm{1/N_{dishes}}$.  
Note that this is the angular power spectrum, and only for a single frequency bin. The signal will be further boosted by 
combining different frequency bins, or computing a 3D power spectrum, noise being uncorrelated between frequency bins, 
while the LSS signal is correlated. However, the presence of foregrounds due to the Galactic synchrotron and the radio sources introduces 
correlation along the frequency axis or redshift, making the 21--cm LSS signal extraction much more challenging. 
}

However, the Tianlai 16-dish array survey should be sensitive enough to detect extra galactic
\HI clumps. The total  21--cm power flux $\Phi_{21}$ received on Earth from an atomic hydrogen clump with mass $M_\mathrm{HI}$, at a luminosity distance $d_L$ can be written as \citep{1998gaas.book.....B}:  
\begin{eqnarray*}
\Phi_{21} 
  & \simeq &   6.4 \times 10^{-20} \frac{M_\mathrm{HI}}{10^9 M_\odot} \, \left( \frac{1 \mathrm{Mpc}}{d_L} \right)^2 
  \hspace{2mm} \mathrm{(W/m^2)}  
\end{eqnarray*}  
Assuming that the clump 21--cm emission frequency dispersion is below 1 MHz ($\Delta v \lesssim 200 \, \mathrm{km/s}$),
we can convert the power flux $\Phi_{21}$ into temperature excess $\Delta T_{21}^\mathrm{pix}$ 
in map pixels covering a solid angle $\delta \Omega \simeq 0.25^2 \, \mathrm{deg^2}$ 
and $\Delta \nu =1$ MHz in frequency:  
\begin{eqnarray*}
P_{21} & = & k_B \, \Delta T_{21}^\mathrm{pix} \, \delta \nu, \hspace{10mm}  k_B = 1.38 \times 10^{-23} \mathrm{J/K} \\
\Delta T_{21}^\mathrm{pix} & = & \Phi_{21} \times \frac{\lambda_{21}^2}{\delta \Omega}  \frac{1}{k \delta \nu} 
\hspace{2mm} \simeq \hspace{1mm} 2.2 \times 10^{20} \, \Phi_{21}  
\end{eqnarray*}  
We can then write the excess temperature $\Delta T_{21}^\mathrm{pix}$  due to \HI clump in 
$\sim 0.25^2 \, \mathrm{deg^2} \times 1 \mathrm{MHz}$ pixels as: 
\begin{eqnarray}
\Delta T_{21} ^\mathrm{pix} & \sim & 14 \times \left( \frac{M_\mathrm{HI}}{10^9 M_\odot} \right) \, \left( \frac{1 \mathrm{Mpc}}{d_L} \right)^2  \hspace{5mm} \mathrm{K}
\end{eqnarray}  

The Tianlai 16-dish polar cap survey should reach a noise level of $\sigma_{\rm noise} \sim 7.5 \, \mathrm{mK}$ for 
map pixels  $\sim 0.25^2 \, \mathrm{deg^2} \times 1 \mathrm{MHz}$. If we consider a 
$3 \sigma \simeq 23 \, \mathrm{mK}$ detection threshold, we see that \HI clumps with masses 
$\sim 5 \times 10^8 M_\odot$ would be detected up to $d_L \lesssim 10 \, \mathrm{Mpc}$ or  
$\sim 5 \times 10^9 M_\odot$ up to $d_L \lesssim 30 \, \mathrm{Mpc}$.
{\rzrefreq 
Based on the HIPASS survey \citep{2005MNRAS.359L..30Z} and ALFALFA survey \citep{2010ApJ...723.1359M}
results, the HI mass function is about $dn/d\ln M_{\rm HI} \sim 10^{-1.4} (\Mpc/h)^{-3}$ and
fairly flat in this mass range, so we estimate that the Tianlai-16 dish array should be able to detect $\sim 10^2 $ such clumps in 
a survey covering $f_{\rm sky} \sim 16\%$ of the sky in the polar cap area. 
} 
\section{Conclusions and outlook}
\label{sec-conclusions}
\rzrefreqa{ A radio instrument with large instantaneous field of view and large bandwidth observing in transit mode can be used 
to perform efficiently a  cosmological neutral hydrogen survey over a significant fraction of the sky. }
A number of such instruments including both single dishes and interferometer arrays are being developed for such surveys. 
For the current generation of instruments,  the number of observables (visibilities) 
already exceeds $10^6$, so reconstructing the sky map from theses observations requires highly efficient algorithms. 

We present our sky map reconstruction method based on the spherical harmonics transformation.
It is shown that the large inverse map reconstruction problem can be decomposed into a set of much smaller independent problems, one for each spherical harmonics $m$-mode, reducing the numerical complexity 
of the problem by several orders of magnitude. 
We have developed an efficient, flexible and parallel code to compute the sky map from transit visibilities.
Our software tools can process visibility data from any transit-type interferometer
performing {\rzrefreq full circle scans} at fixed declinations.  

In this paper we focus on the case of dish arrays and study  several instrument configurations and scan strategies. 
We have computed mock visibility time streams then reconstructed sky maps from these visibilities.
The instrument response matrices, the transfer function and the noise covariance matrices have been computed
for the different configurations.

First, the relatively simple cases of arrays with 4 dishes are investigated, including the 
PAON-4 test interferometer at Nancy, France which has a triangular layout and several 
regular $2 \times 2$ layouts. By considering these examples, we study the impact of the array 
configuration and the survey strategy on the sky reconstruction performance. 
We show in particular the importance of having short baselines 
and a compact array layout to obtain a complete coverage of the $(\ell, m)$ space without {\it holes}.
We show also that the uniformity of the response function over the $(\ell, m)$ space is obtained 
by having a large number of independent baselines. 

The analysis is then extended to the case of the Tianlai 16-dish pathfinder array located in 
Hongliuxia, Balikun County, Xinjiang, China. We consider two compact layouts to achieve 
good $(\ell,m)$ space coverage: the regular array which is a $4\times 4$ square 
and a circular array with one antenna in the centre and 15 dishes distributed in two concentric rings. 
The Tianlai array configuration results from a compromise between 
the $(\ell,m)$ coverage, and minimising the blocking factor for observations far from the zenith. 
The transfer function and noise angular power spectrum are computed for these configurations in
mid-latitude observations \rzrefreqa{as well as for a polar cap survey with the circular configuration. }
We show that  the circular configuration provides a more uniform 
coverage of the $(\ell,m)$ space and yields better results than the regular $4\times 4$ array,
in terms of response uniformity, angular resolution, beam symmetry and noise.  
It should be noted that for other issues such as the calibration, the redundant baselines 
in the $4\times 4$ array may be an advantage. 

Although the Tianlai 16-dish pathfinder array is too small to reach the sensitivities necessary to observe 
the cosmological \HI signal, it is able to make  good measurement of the Galactic \HI signal, reconstructing  
sky modes and \HI power spectrum up to $\ell < 1000$ at 1420 MHz. It would also be able to detect 
extra Galactic \HI clumps with masses $\sim 10^9 M_\odot$ up to $\sim 30$ Mpc,
\rzrefreqa{in particular with the polar cap survey which reaches higher sensitivity at the expense of reduced 
survey area. } 
 
This is the first of a series of papers on our transit array data processing method. 
The basic formalism has been presented and we explored some of the features of transit array 
observation and map making, {\rzrefreq taking the PAON-4 and Tianlai 16-dish arrays as specific examples.
We have made a number of simplifying assumptions and ignored some complications such as the polarisation. 
Indeed, as shown by \citet{2015PhRvD..91h3514S}, the method can easily be extended to handle polarised sky emissions. 
The question of array calibration, impact of calibration uncertainties, imperfect knowledge of the single feed 
response or array geometry will be addressed in future work as well as the extension to polarisation and 
foreground subtraction. }


\section*{Acknowledgements}
We would like to thank Shifty Zuo, Yichao Li, Ue-Li Pen and Richard Shaw for discussions. 
The Tianlai project is supported by the MoST 863 programme grant 2012AA121701 and the CAS Repair and Procurement grant. 
PAON-4 project is supported by PNCG,  Observatoire de Paris, Irfu/CEA and LAL/CNRS. 
J.~Z. was supported by China Scholarship Council. X.~C. is supported by the CAS strategic Priority Research Programme XDB09020301, 
and NSFC grant 11373030. F.~W. is supported by NSFC grant 11473044.

\bibliographystyle{mnras}
\bibliography{jmapmaking}
\label{lastpage}

\appendix

\section{Map making tools}
\label{sec-mapmakingcode}
\subsection{The software structure}

The map reconstruction software and associated tools ({\bf JSkyMap}) are written in C++ and use 
the SOPHYA (SOftware for PHYsics Analysis) \footnote{\url{ http://www.sophya.org}} class library. SOPHYA is a collection of C++ classes designed to ease data analysis software development and provide the following services to the map making software:
\begin{itemize}
\item[-] Input/Output services in different formats, including ASCII, FITS and the native PPF SOPHYA binary format
\item[-] Various standard numerical analysis algorithms, including FFT and linear algebra and interface to LAPACK
\item[-] Several map pixelisation in spherical geometry, including the HEALPix format \cite{2005ApJ...622..759G} 
\item[-] Spherical Harmonics Transform 
\item[-] Classes to perform parallel computation 
\end{itemize}

We have developed the sky map reconstruction code both in rectangular geometry ($\uv$ plane) and spherical 
geometry ($(\ell,m)$ plane). The rectangular geometry can be used when observing a narrow band in declination,
at low declinations. We have performed a number of cross checks, for example computing visibilities in spherical
geometry and performing the reconstruction in rectangular geometry. The code has similar structure in the two 
geometry. For the sake of clarity, we present here only the map reconstruction code in spherical geometry.
The software is organised around few main classes:
\begin{itemize}
\item[-] {\tt BeamTP} and {\tt BeamLM} classes which represent the beam for a single feed or a pair of feeds in angular $(\vec{\hat{n}}=(\theta,\varphi))$ domain in spherical geometry and in the spherical Harmonics coefficient domain  ($(\ell, m)$ plane). The {\tt BeamVis} class computes the $(\ell, m)$ plane response for a parir of feeds/antenna given the
the baseline, and the array position in latitude. 
\item[-] The template class {\tt PseudoInverse} provides the specific services to handle the computation 
of $\mathbf{H}_m$ matrices and the noise covariance matrices.
\item[-] Utility classes and function to handle the computation of the set of baselines from the antenna positions in an array.
\item[-]  The  {\tt JSphSkyMap } is the main class in the reconstruction code.
It computes the $\mathbf{L}_m$ and $\mathbf{H}_m$ matrices, starting from a set of beams in the $(\ell, m)$ 
plane corresponding to an instrument layout and sky scanning strategy.
It provides also methods to computes mock visibility data, given an input sky, as well as methods to reconstruct 
the sky from visibilities.  This class implements parallelism at the level of 
$\mathbf{L}_m$ and $\mathbf{H}_m$ matrices for different $\ell$. 
\end{itemize}

\begin{figure}
\includegraphics[width=0.5\textwidth]{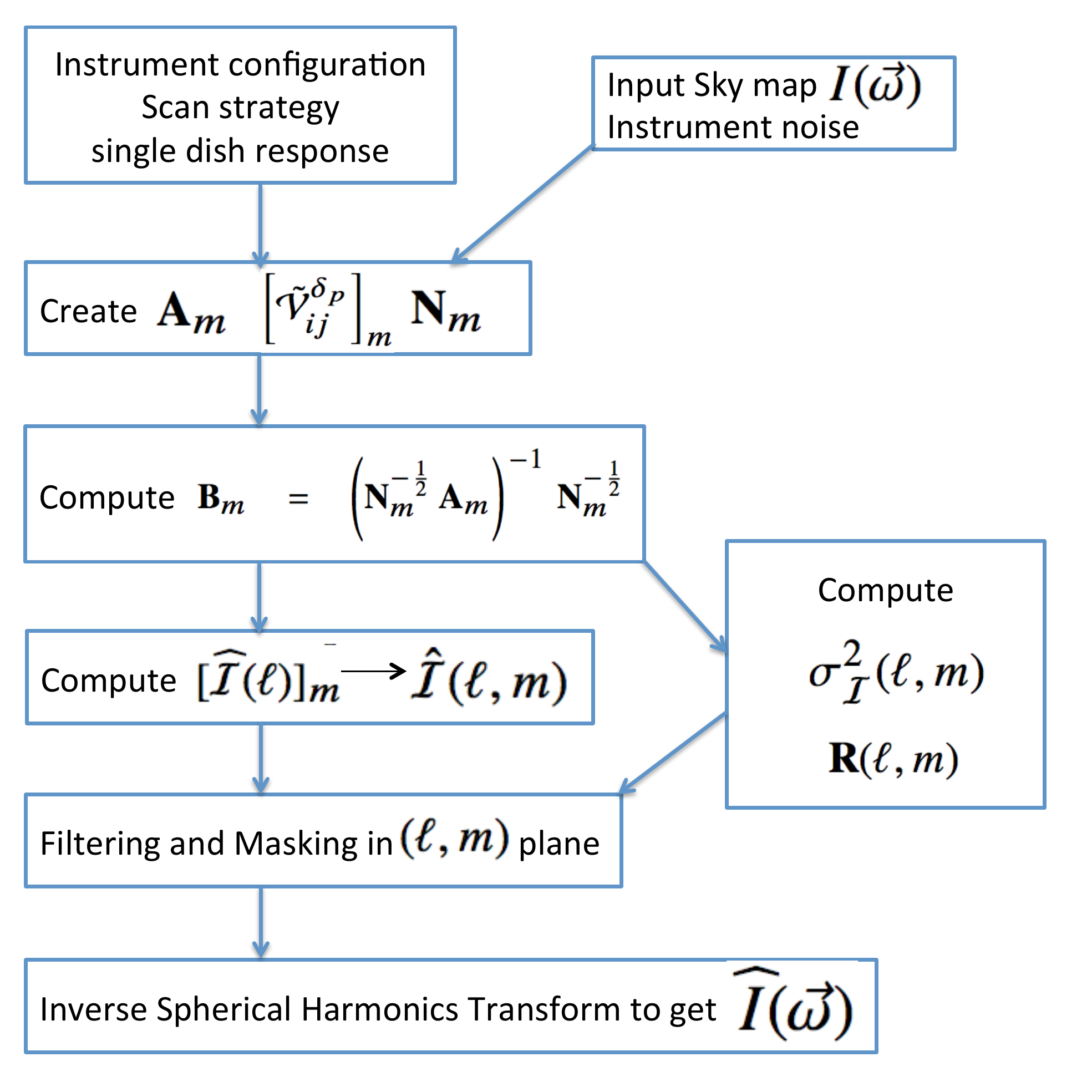}
\caption{Block Diagram for the map-reconstruction code}
\label{fig-code}
\end{figure}

The overall functional  structure of the {\bf JSkyMap} code is shown in Fig.\ref{fig-code}.
The major computation steps are listed below:
\begin{enumerate}
\item We compute first the baselines from the array configuration, i.e. the coordinates of the array elements. 
To simplify numerical  handling,  we replace redundant baselines by a single beam, scaling the noise level accordingly
$\propto 1/n_{rb}$, where  $n_{rb}$ is the number of antenna pairs with the exact same baseline. 
\item We compute then the beams in the $(\ell,m)$ plane from the baselines and scan strategy (the observed declinations on sky). As the beam computation involves Spherical Harmonics Transform (SHT) which is computation 
intensive, multi-threaded computation has been implemented for this step.  
\item One can then compute the $\mathbf{L}_m$ and the visibilities noise covariance matrices for each $m$-mode 
and then the corresponding $\mathbf{H}_m$ matrix, using the  pseudo-inverse computation. The computation for theses 
steps benefits also from multo-threaded implementation, taking advantage of $m$-mode parallelism. 
\item Mock visibilities (with or without noise) can be computed using the $\mathbf{L}_m$  matrices.
\item Finally $\mathbf{H}_m$ matrices can be used to compute estimated  Sky spherical harmonics 
coefficients from mock or observed visibilities, as well as the corresponding error covariance matrix. 
The sky map is recovered using an inverse SHT.
\end{enumerate}


\subsection{Pseudo-inverse matrix} 
\label{sec-pseudoinverse}

The singular value decomposition (SVD) provides an efficient way for computing the pseudo-inverse of a matrix. 
An $m \times n$ real or complex matrix A can be factorized in the form 
$$ A = U \Sigma V^\dagger ,$$
where U is an $m \times m$ real or complex unitary matrix, $\Sigma$ is an $m \times n$  rectangular diagonal matrix, i.e. all non-diagonal elements are zero, with nonnegative real numbers on the diagonal, and $V^\dagger$ (the conjugate transpose of $V$, or simply the transpose 
of $V$ if $V$ is real) is an $n\times n$ real or complex unitary matrix.
The diagonal entries $\Sigma_{ii}$ of $\Sigma$ are known as the singular values of A. A common convention is to list the singular values in descending order, then the diagonal matrix $\Sigma$ is uniquely determined by A, though the matrices U and V are not unique.

Using SVD, the pseudo-inverse of the matrix A  is given by
\begin{eqnarray}
\label{b1}
B \equiv A^{-1} = V \bar{\Sigma}^{-1} U^\dagger
\end{eqnarray}
where $\bar{\Sigma}^{-1}$ is formed by replacing every nonzero diagonal entry $\Sigma_{i,i}$ by its reciprocal $1/\Sigma_{i,i}$ and transposing the resulting matrix. 

Due to limited numerical precision in the computation, even zero elements of $\Sigma_{i,i}$ will have some small non-zero value, 
which would give rise to large $\Sigma_{i,i}^{-1}$ and affect the result greatly if left unattended. To ensure the stability of the computation, 
the small eigenvalues of $\Sigma$ are set to 0 before the inversion, and its inverse also set to 0 and ignored in subsequent computation. In practice, we set some threshold value for the diagonal elements. If the diagonal element $\Sigma_{i,i} < \Sigma_{0,0} \times \epsilon_r$ or 
$\Sigma_{i,i} < \epsilon_a$,we set $\Sigma_{i,i} \to 0$, where $\Sigma_{0,0}$ is the largest eigenvalue, and
$\epsilon_r, \epsilon_a$ are small threshold values for the relative and 
absolute size of the eigenvalues. We experimented with different values of $\epsilon_r, \epsilon_a$, finally choosing $\epsilon_r =0.02$
and $\epsilon_a=0.01$, below which we often run into stability problems in the computation.

The standard solution of the measurement equation Eq.(\ref{eq-visibility}) is given in Eq.(\ref{eq-inverse}), but this form of solution
involves several products of large non-diagonal matrices. For the  Tianlai 16-dish array, the A matrix size is $\sim (3000,1500)$ for each 
$m$-mode,  which takes a lot of time and computing resources.
If $\mathbf{N}_m$ matrix is diagonal and positive, the computation could be further simplified. 
For a symmetric or hermitic matrix $N$, the SVD decomposition leads to $N^{-1}=U \Sigma U^\dagger $.
For a symmetric matrix, $U$ is real and $U^\dagger = U^T$ where $^T$ denotes transpose operation.  
We can then define $N^{-1/2}\equiv \Sigma^{-1/2} U^\dagger$, where $\Sigma^{-1/2}$ is obtained simply by taking the inverse 
of square root of each non-zero diagonal element of $\Sigma$. The $N$ matrix can then be factorised  
as $N^{-1} = \left( N^{-\frac{1}{2}} \right)^\dagger N^{-\frac{1}{2}}$, and
substitute this into Eq.(\ref{eq-visibility}), we have
\begin{eqnarray}
\label{eq-invsqrt}
B &=&  (A^\dagger N^{-1} A)^{-1} A^\dagger N^{-1} \nonumber \\
&=& [A^\dagger (N^{-1/2})^\dagger N^{-1/2} A ]^{-1} A^\dagger  (N^{-1/2})^\dagger N^{-1/2}   \nonumber  \\
&=& (N^{-1/2} A)^{-1} [A^\dagger (N^{-1/2})^\dagger]^{-1}  [A^\dagger  (N^{-1/2})^\dagger] N^{-1/2}   \nonumber  \\
&=& (N^{-\frac{1}{2}} A)^{-1} N^{-\frac{1}{2}}  
\end{eqnarray}
where we derived the fourth line from the third line using  $(\mathbf{X} \mathbf{Y})^{-1}=Y^{-1} X^{-1}$ for matrix $\mathbf{X,Y}$,
which holds for pseudo-inverse as well as the usual matrix inverse. 
If $N$ is diagonal as the case of this paper, the final expression Eq.(\ref{eq-invsqrt}) can be computed much more 
easily, as $N^{-1/2}$ can be computed directly ($U=I$), and 
the $A$ matrix is only multiplied by the diagonal matrix $N^{-1/2}$. 

Table \ref{table-SVD} shows a few examples of the SVD computation for the pseudo-inverse $(\mathbf{N}_m^{-1/2} \mathbf{L}_m)^{-1/2}$
matrix for the PAON4 configuration. In this table we show for the selected $m$ mode the largest eigenvalue, the 
threshold value,  and the number of non-zero eigenvalues, i.e. the number of eigenvalues which are above the threshold (
eigenvalues below threshold are set to zero to avoid contamination by numerical error).   
The threshold is different for each case, because the largest eigenvalue is different, and we see that among the examples listed
here the threshold values change, implying that most are determined by the relative criterion, i.e. that the eigenvalue must be
less than 0.02 of the largest eigenvalue. As $m$ increases, the non-zero modes decreases, eventually reaching 0.01, which is the absolute
threshold adopted here. We have also tried to vary the threshold a bit, such variations affect the number of non-zero modes, but not by 
much as long as we try to keep the numerical solution stable.  

\begin{table}
\caption{The SVD computation for a few $m$-modes in the PAON-4 case. }
\label{table-SVD}
\begin{tabular}{|c|c|c|c|} \hline
m modes  &  frist eigenvalue & thresholds & non-zero eigenvalues  \\ \hline
1  &  5.15878  &  0.103176  & 53 \\ \hline
50  &  1.93873  &  0.0387746  & 71  \\ \hline
100  &  1.71407  &  0.0342814  & 51  \\ \hline
150  &  1.38509  &  0.0277018  & 45  \\ \hline
200  &  1.56662  &  0.0313324  & 41  \\ \hline
250  &  1.39035  &  0.0278069  & 28  \\ \hline
300  &  0.604846  &  0.0120969  & 16  \\ \hline
350  &  0.0449966  &  0.01  & 4  \\ \hline
\end{tabular}
\end{table}


\section{Extension to polarisation}
\label{sec-polarisedmapmaking}
The formalism presented in section \ref{sec-transitmapmaking} deals with unpolarised sky signal which is suitable for the description of the cosmological 21-cm emission. However, it is well known that most RFI are strongly polarised, as well  
as components of the foregrounds emission, in particular emission from compact radio sources \citep{2014ApJS..212...15F} or 
the synchrotron emission of our own galaxy. For polarised emission, the Faraday rotation due to the interstellar magnetic field imprints frequency dependent structures on the polarised emission, increasing foreground separation difficulty. 
The reconstruction of the polarised sky emission maps is thus mandatory for intensity mapping project.  
In this section we describe briefly how the method described in this paper and the corresponding code can be extended to 
handle reconstruction of polarised brightness maps from polarised visibility signals. We largely follow the results given in the reference \citep{2015PhRvD..91h3514S}. 

Polarisation characterises the vectorial nature of electromagnetic (EM) radiation, representing a fundamental property separate from
its frequency and intensity. The polarisation of an antenna refers to the orientation of the electric field of the radio wave with 
respect to the Earth's surface and is determined by the physical structure of the antenna and by its orientation. 
We assume that each antenna is equipped with dual polarisation receivers, measuring two orthogonal linear polarisations 
$(\hat{x},\hat{y})$ of the incoming electromagnetic field \citep{2004A&A...420..437C}, for example 
a component $\hat{x}$ parallel to the horizontal plane $\hat{y}$ parallel to the meridian plane. The measured electric signal for each 
polarisation is a combination of the corresponding projection of the electric field contributions coming from different and incoherent directions of the sky. 
The polarization state of electromagnetic waves is often described using a 4-element column vector corresponding to the Stokes parameters $S=(I,Q,U,V)^T$ where superscript T denotes the matrix transpose. If $e_x$ and $e_y$ denotes the two electric field components transverse to the line of sight, one gets 
\begin{align}
I &= \phantom{-i} \langle e_x e^\ast_x\rangle + \langle e_y e^\ast_y\rangle&
Q &= \phantom{-i} \langle e_x e^\ast_x\rangle - \langle e_y e^\ast_y\rangle \nonumber \\
U &= \phantom{-i} \langle e_x e^\ast_y\rangle + \langle e_y e^\ast_x\rangle&
V &= -i (\langle e_x e^\ast_y\rangle - \langle e_y e^\ast_x\rangle) 
\label{eq-stokes}
\end{align}
where the $\langle \rangle$ denotes a time average, and we have omitted the direction $(\vec{\hat{n}})$ 
dependence for simplicity. The visibilities $\mathcal{V}_{p_i, p_j}$  have to be computed for all signal pairs, 
$(p_i , p_j)$ indices identifying the antenna pair $(i,j)$, as well as the polarisation probe $x$ or $y$. 
The full set of visibilities $\mathcal{V}_{p_i, p_j}$ can be split in two sets:  $x$ and $y$ polarisations auto 
and cross correlations $\mathcal{V}_{ij}^{xx}, \mathcal{V}_{ij}^{yy}$ and cross polarisation
visibilities $\mathcal{V}_{ij}^{xy}, \mathcal{V}_{ij}^{yx}$. For an array with N dual polarisation receivers, 
there will be a total of $2 N^2$ visibilities, corresponding to $2 N$ autocorrelations for the $x$ and $y$ polarisation 
signals, $\frac{N \, (N-1) }{2} $ cross correlations visibilities for each of $xx$ and $yy$ polarisation signal pairs, and  
$\frac{N^2}{2}$ visibilities for each of the cross polarisation $xy$ and $yx$ pairs. 
\begin{align}
\mathcal{V}_{p_i, p_j}  & = & \left[  \mathcal{V}_{ij}^{xx}  ;  \mathcal{V}_{ij}^{yy}  ;  \mathcal{V}_{ij}^{xy}  ;  \mathcal{V}_{ij}^{yx} \right]  \\
p_i & = & \left\{ (i,x), (i,y) \right\} \hspace{10mm} p_j  =  \left\{ (j,x), (j,y) \right\}
\end{align}
The generalization of \RefEq{eq-visib}\ reads
\begin{equation}
\mathcal{V}_{p_i p_j} = \iint \, \sum_a \mathbf{L}^a_{p_i p_j}(\vec{\hat{n}}) S_a(\vec{\hat{n}})  \, d \vec{\hat{n}}  
\label{eq-visib-stokes}
\end{equation}
where the sum on index $a$ runs over the four Stokes parameters. The four  beams 
$$ \mathbf{L}_{p_i p_j} =  \left\{ L_{p_i p_j}^I, L_{p_i p_j}^Q, L_{p_i p_j}^U, L_{p_i p_j}^V \right\} $$
are a generalisation of the beam pattern of Eq.(\ref{eq-beampattern}) that takes into account the 
response of each of the two linear polarisation
probes of the feed, including the response to the incoming electric field signal, 
as well as all possible leakage sources from one polarisation to the other. 
It might include other effects impacting polarisation measurement, such as polarisation direction rotation 
or leakage due to the atmosphere and/or earth magnetic field.  
The Stokes parameters decomposition requires spin-weighted spherical harmonics \citep{1997PhRvD..55.1830Z} 
\rzcheckdone{(Il y a une coquille dans l'appendix de cet article pour le conjugue de sYlm)} 
with spin-0 for $I$ and $V$ and spin-2 (${}_sY_{\ell m}$) for $Q$ and $U$. 
From the two real quantities $U$ and $Q$, we define two complex linear combinations, and corresponding 
spherical harmonics coefficients: 
$\bar{Q} = (Q+iU)/2$ and $\bar{U}= (Q-iU)/2$:
\begin{align}
I(\vec{\hat{n}}) &\xrightarrow{{}_0Y_{\ell m}} \mathcal{I}_{\ell m}& 
\bar{Q}(\vec{\hat{n}}) &\xrightarrow{{}_{+2}Y_{\ell m}} \mathcal{\bar{Q}}_{\ell m} \nonumber \\
\bar{U}(\vec{\hat{n}}) &\xrightarrow{{}_{-2}Y_{\ell m}} \mathcal{\bar{U}}_{\ell m}&
V(\vec{\hat{n}}) &\xrightarrow{{}_0Y_{\ell m}} \mathcal{V}_{\ell m}
\label{eq-stokes-alm}
\end{align}
The angular responses of the polarised beam $\mathbf{L}$ may also be decomposed  in spherical harmonics.
As for the Stokes parameters, we define linear combination of $Q$ and $U$ beams:
$L^{\bar{Q}}= L^Q- i L^U$ and $L^{\bar{U}}=L^Q + i L^U$ to match the definition of $\bar{Q}$ and $\bar{U}$ respectively.
\begin{align}
L_{p_i p_j}^I &\xrightarrow{{}_0Y_{\ell m}} \mathcal{L}_{p_i p_j;\ell m}&
L_{p_i p_j}^{\bar{Q}} &\xrightarrow{{}_{-2}Y_{\ell m}}  \mathcal{L}_{p_i p_j;\ell m}^{\bar{Q}}& \nonumber \\
L_{p_i p_j}^{\bar{U}} &\xrightarrow{{}_{+2}Y_{\ell m}}  \mathcal{L}_{p_i p_j;\ell m}^{\bar{U}}&
L_{p_i p_j}^V &\xrightarrow{{}_0Y_{\ell m}} \mathcal{L}_{p_i p_j;\ell m}^V
\label{eq-stokes-beam-alm}
\end{align}
Using the symmetry property of spin-2 spherical harmonics ${}_sY^\ast_{l,m} = (-1)^{s+m}{}_{-s}Y_{l,-m}$,
and the orthogonality of the spin-weighted spherical harmonics, the extension of Eq.(\ref{eq-vis-alm})  reads
\begin{multline}
\mathcal{V}_{p_i p_j} = \sum_{ml} (-1)^m \left(  \mathcal{L}_{p_i p_j;\ell,-m}^I \mathcal{I}_{\ell m} 
+  \mathcal{L}_{p_i p_j;\ell,-m}^{\bar{Q}} \mathcal{\bar{Q}}_{\ell m} \right. \\ \left.
+  \mathcal{L}_{p_i p_j;\ell,-m}^{\bar{U}} \mathcal{\bar{U}}_{\ell m}
+  \mathcal{L}_{p_i p_j;\ell,-m}^V \mathcal{V}_{\ell m}
\right)
\end{multline}
It is convenient to decompose  $ \mathcal{\bar{Q}}$ and  $ \mathcal{\bar{U}}$  with 
the \emph{gradient} ($E$) and \emph{curl} ($B$) components and the corresponding beams $\mathcal{L}_{p_i p_j}$  
\begin{align}
\mathcal{\bar{Q}}_{\ell m} &= - \mathcal{E}_{\ell m} - i\mathcal{B}_{\ell m} &  
 \mathcal{L}_{p_i p_j;\ell m}^{\bar{Q}} &= \left(- \mathcal{L}_{p_i p_j;\ell m}^{E} + i  \mathcal{L}_{p_i p_j;\ell m}^{B}\right)/2 
\nonumber \\ 
 \mathcal{\bar{U}}_{\ell m}  &= -\mathcal{E}_{\ell m} +i \mathcal{B}_{\ell m}&  
  \mathcal{L}_{p_i p_j;\ell m}^{\bar{U}} &= \left(- \mathcal{L}_{p_i p_j;\ell m}^{E} - i  \mathcal{L}_{p_i p_j;\ell m}^{B}\right)/2
\end{align}  
Then, 
\begin{multline}
\mathcal{V}_{p_i p_j} = \sum_{m \ell} (-1)^m \left(  \mathcal{L}_{p_i p_j;\ell,-m}^I \mathcal{I}_{\ell m} 
+  \mathcal{L}_{p_i p_j;l,-m}^{E} \mathcal{E}_{\ell m} \right. \\ \left.
+  \mathcal{L}_{p_i p_j;l,-m}^{B} \mathcal{B}_{\ell m}
+  \mathcal{L}_{p_i p_j;l,-m}^V \mathcal{V}_{\ell m}
\right)
\end{multline}
As all Stokes parameters are real functions then $\bar{U}^\ast(\vec{\hat{n}}) = \bar{Q}(\vec{\hat{n}})$ and this leads the relation in harmonic space $\bar{Q}_{\ell m} = (-1)^m \bar{U}^\ast_{l,-m}$ and to relations which extend the case of $\mathcal{I}_{\ell m}$ as 
\begin{equation}
X_{l,-m} = (-1)^m\ X^\ast_{\ell m} \quad \mathrm{X} \in \{\mathcal{I},\ \mathcal{E},\ \mathcal{B},\ \mathcal{V}\}
\end{equation}

So, one can extend both the Fourier decomposition Eq.~(\ref{eq-vis-I-FFT}) as well as the positive and negative m-mode separation 
Eqs.~(\ref{eq-mmode1}), (\ref{eq-mmode2}). 
\begin{align}
\tilde{\mathcal{V}}_{p_i p_j}(m) & = \sum_{\ell=|m|}^{+\ell_{max}} \sum_\mathcal{X} (-1)^m \mathcal{L}^{X}_{p_i p_j;l,-m} \mathcal{X}_{\ell m} \\
\tilde{\mathcal{V}}_{p_i p_j}^\ast(-m) & = \sum_{\ell=|m|}^{+\ell_{max}} \sum_\mathcal{X} \mathcal{L}^{X\ast}_{p_i p_j;l,m} \mathcal{X}_{\ell m}
\end{align}
with $\mathcal{X}=\mathcal{I},\mathcal{E},\mathcal{B},\mathcal{V}$.

Extending our map making software to perform computation for the polarised case would be rather straightforward, except maybe 
for the computation of the polarised beams  $\left\{ L_{p_i p_j}^I, L_{p_i p_j}^Q, L_{p_i p_j}^U, L_{p_i p_j}^V \right\}$, from 
individual feed polarised beam responses. The implementation of the extension is postponed to future work.

%
\end{document}